\documentclass[aps,pra,twocolumn,superscriptaddress,linenumbers%
 reprint,
 amsmath,amssymb,
 aps,
prl,
]{revtex4-2}

\usepackage{graphicx}
\usepackage{dcolumn}
\usepackage{bm}
\usepackage{xcolor}

\usepackage{hyperref}
\hypersetup{
   pdfpagemode=None,
   pdfstartpage=1,
   pdfmenubar=true,
   pdftoolbar=true,
   colorlinks = true,
   linkcolor=blue,
   citecolor=blue,
   urlcolor=blue,
   bookmarksopen=false
 }

\usepackage{physics}
\usepackage{comment}

\newcommand{\Cite}{\unskip~\cite}
\newcommand{\Eqref}{\unskip~\eqref}
\newcommand{\figref}{\unskip~\ref}
\newcommand{\qo}[1]{``#1''} 

\usepackage{csquotes}
\begin{document}

\preprint{APS/123-QED}

\title{Tomographic characterization of non-Hermitian Hamiltonians in reciprocal space}


\author{Francesco Di Colandrea}
\email{francesco.dicolandrea@unina.it}
\affiliation{%
 Dipartimento di Fisica \qo{Ettore Pancini}, Universit\`{a} degli Studi di Napoli Federico II, 
 Complesso Universitario di Monte Sant'Angelo, Via Cintia, 80126 Napoli, Italy
}%

\author{Fabrizio Pavan}
\thanks{FDC and FP contributed equally to this work.} 
\affiliation{%
 Dipartimento di Fisica \qo{Ettore Pancini}, Universit\`{a} degli Studi di Napoli Federico II, 
 Complesso Universitario di Monte Sant'Angelo, Via Cintia, 80126 Napoli, Italy
}%

\author{Sarvesh Bansal}
\affiliation{%
Dipartimento di Fisica \qo{Ettore Pancini}, Universit\`{a} degli Studi di Napoli Federico II, Complesso Universitario di Monte Sant'Angelo, Via Cintia, 80126 Napoli, Italy
}%

\author{Paola Savarese}%
\affiliation{%
Dipartimento di Fisica \qo{Ettore Pancini}, Universit\`{a} degli Studi di Napoli Federico II, Complesso Universitario di Monte Sant'Angelo, Via Cintia, 80126 Napoli, Italy
}%

\author{Grazia Di Bello}%
\affiliation{%
Dipartimento di Fisica \qo{Ettore Pancini}, Universit\`{a} degli Studi di Napoli Federico II, Complesso Universitario di Monte Sant'Angelo, Via Cintia, 80126 Napoli, Italy
}%
\affiliation{INFN, Sezione di Napoli - Complesso Universitario di Monte Sant'Angelo, Via Cintia, 80126 Napoli, Italy}%

\author{Giulio De Filippis}%
\affiliation{INFN, Sezione di Napoli - Complesso Universitario di Monte Sant'Angelo, Via Cintia, 80126 Napoli, Italy}%
\affiliation{%
SPIN-CNR and Dipartimento di Fisica \qo{Ettore Pancini} - Università di Napoli Federico II, 80126 Napoli, Italy
}%

\author{Carmine Antonio Perroni}%
\affiliation{%
INFN, Sezione di Napoli - Complesso Universitario di Monte Sant'Angelo, Via Cintia, 80126 Napoli, Italy}%
\affiliation{%
SPIN-CNR and Dipartimento di Fisica \qo{Ettore Pancini} - Università di Napoli Federico II, 80126 Napoli, Italy
}%

\author{Donato Farina}%
\email{donato.farina@unina.it}
\affiliation{%
Dipartimento di Fisica \qo{Ettore Pancini}, Universit\`{a} degli Studi di Napoli Federico II, Complesso Universitario di Monte Sant'Angelo, Via Cintia, 80126 Napoli, Italy
}%

\author{Filippo Cardano}%
\affiliation{%
Dipartimento di Fisica \qo{Ettore Pancini}, Universit\`{a} degli Studi di Napoli Federico II, Complesso Universitario di Monte Sant'Angelo, Via Cintia, 80126 Napoli, Italy
}%

\begin{abstract}
Non-Hermitian Hamiltonians enrich quantum physics by extending conventional phase diagrams, enabling novel topological phenomena, and realizing exceptional points with potential applications in quantum sensing.
Here, we present an experimental photonic platform capable of simulating a non-unitary quantum walk generated by a peculiar type of non-Hermitian Hamiltonian, largely unexplored in the literature.
The novelty of this platform lies in its direct access to the reciprocal space, which enables us to scan the quasi-momentum across the entire Brillouin zone and thus achieve a precise tomographic reconstruction of the underlying non-Hermitian Hamiltonian, indicated by the comparison between theoretical predictions and experimental measurements.
From the inferred Hamiltonian, it is possible to retrieve complex-valued band structures, resolve exceptional points in momentum space, and detect the associated parity-time symmetry breaking through eigenvector coalescence.
Our results, presented entirely in quasi-momentum space, represent a substantial shift in perspective in the study of non-Hermitian phenomena.
\color{black}
\end{abstract}


\maketitle

{\it Introduction.--- }%
Non-Hermitian physics \Cite{ashida2020non,el2018non} has emerged as a powerful framework to describe losses and gains in quantum systems, with broad implications in topological condensed matter\,\cite{PhysRevX.8.031079, PhysRevX.9.041015, PhysRevA.97.052115, PhysRevB.95.201407,PhysRevLett.102.065703, PhysRevLett.132.156901,PhysRevB.107.L121101}, skin effects \Cite{PhysRevLett.125.186802, PhysRevLett.123.170401}, and enhanced quantum sensing \Cite{PhysRevLett.133.180801}.
In this framework, the system dynamics is governed by an effective non-Hermitian Hamiltonian 
\begin{equation}
    \mathcal{H}_{\mathrm{eff}} = H + i\Delta ,
\end{equation}
where $H=H^\dagger$ is the Hermitian component and ${\Delta=\Delta^\dagger}$ is a Hermitian operator encoding the non-Hermitian contribution (gain–loss term).
Non-Hermiticity generally leads to complex spectra. 
This effective description can be formally derived by post-selecting 
the no-quantum-jump trajectories in an open-system dynamics, 
leading to a non-Hermitian Hamiltonian governing the conditional evolution 
\cite{PhysRevA.100.062131, roccati2022non}.
Noncommutativity of operators is a hallmark of quantum mechanics, 
setting it apart from classical physics, with the Heisenberg uncertainty 
relation being its most striking consequence. 
In the non-Hermitian framework, if the Hermitian and anti-Hermitian components commute, $[H,\Delta]=0$, the eigenvectors remain orthogonal. By contrast, when 
\begin{equation}
    [H,\Delta] \neq 0 ,
\end{equation}
the eigenvectors generally become non-orthogonal and may coalesce at critical parameter values, a defining feature of exceptional points \Cite{doppler2016dynamically}. Concretely engineering non-commuting gain-loss terms $\Delta$ is of paramount importance to advance quantum sensing and topological matter. At an exceptional point, the eigenvectors coalesce, and the geometric multiplicity drops below the algebraic multiplicity. Near an exceptional point, small perturbations can induce disproportionately large changes in the system spectral response, providing a powerful mechanism for ultra-sensitive detection \Cite{eps-sensors,eps-science}. This property has inspired a new generation of quantum sensors capable of detecting extremely weak signals, such as minute variations in magnetic fields \Cite{FAN2024110434}, refractive indices \Cite{PhysRevLett.112.203901}, or mechanical \Cite{Wu2022-wx}. In some architectures, higher-order exceptional points, where multiple modes simultaneously coalesce, are predicted to yield even greater sensitivity enhancements \Cite{PhysRevLett.123.180501}.

In modern quantum technologies and precision sensing, parity–time (PT) symmetry and exceptional points have emerged as pivotal concepts, enabling systems to transcend the limits of traditional Hermitian physics by exploiting the interplay between gain and loss \Cite{Nori-PT,Longhi-PT}. In PT-symmetric systems, a delicate balance between these two effects can produce entirely real energy spectra up to a critical threshold, beyond which the symmetry breaks and the eigenvalues become purely complex.

In this Letter, we present a photonic platform capable of simulating a non-Hermitian Hamiltonian that differs from those derived from the Hatano–Nelson model \Cite{Hn1,Hn2}. In that class of models, non-Hermiticity typically arises from non-reciprocal hopping amplitudes \Cite{Zhang2021,PhysRevA.97.052115}. By following this approach, for instance, a non-Hermitian SSH model can be obtained by introducing different intra-cell hopping amplitudes for leftward and rightward propagation along the chain. In our case, instead, the model features reciprocal hopping with identical amplitudes, while the non-Hermitian character originates from the left and right inter-cell hopping terms, which are not complex conjugates of each other \Cite{PhysRevResearch.6.023202}. Additionally, the model includes a complex on-site potential, introducing a finite probability for the particle to remain localized within a given cell. A distinctive feature of our platform is the ability to resolve and scan the quasi-momentum in reciprocal space. This direct reciprocal-space access allows us to probe the full Brillouin Zone (BZ) and perform an accurate tomographic reconstruction of the eigenvectors of the Hamiltonian. Such flexibility is not available, for example, in transmon-qubit \Cite{naghiloo2019quantum} or trapped-ion \Cite{lu2025dynamical} implementations, and constitutes a central novelty of the present work.

This work experimentally investigates the interplay between topology, sublattice, and PT symmetry \Cite{bender1998real, RevModPhys.96.045002, PhysRevLett.103.093902}, by scanning the system in reciprocal space, using a photonic quantum-walk platform that can be tuned to realize a family of effective Hamiltonians within the same setup. By means of a tomographic protocol, we can establish the properties of the non-Hermitian model implemented in our platform. We show that the Hamiltonian admits a well-defined topological characterization, described by a discontinuous jump in the winding number, which enables us to reconstruct the phase diagram. The physics is further enriched by the appearance of exceptional points along the branch line where the transition occurs, at specific critical quasi-momentum values. Finally, we show that our Hamiltonian exhibits spontaneous PT-symmetry breaking \Cite{bender1998real, RevModPhys.96.045002, PhysRevLett.103.093902}.
\color{black}

\begin{figure}[t!]
    \centering
    \includegraphics[width=\columnwidth]{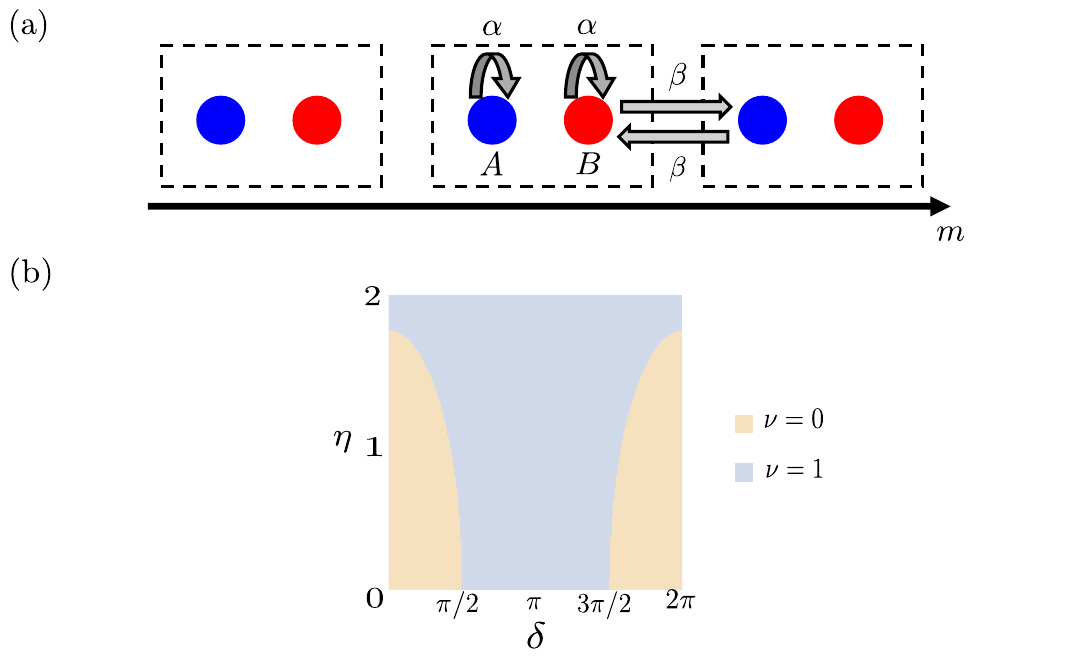}
    \caption{\textbf{Non-Hermitian topological quantum walk.} (a)~At each time step, under the action of the operator $T$ (see Eq. \Eqref{eqn:translation}), the walker maintains the state with probability $\abs{\alpha}^2$, while coupling to neighboring sites with equal probability $\abs{\beta}^2$. The coupling coefficients to the left and right sites are complex, but not related by complex conjugation, which makes the evolution non-unitary. (b)~Topological phase diagram as a function of the parameters $\delta$ and $\eta$ (see Eq. \Eqref{eqn:mapping}).   
    }
    \label{fig:fig1}
\end{figure}

{\it Effective Hamiltonian.---} Quantum walks (QWs) serve as a paradigmatic model for simulating complex lattice Hamiltonians \Cite{Venegas_Andraca_2012,cardanoscience}, probing quantum transport \Cite{MARES2020126302} and topological phenomena \Cite{PhysRevA.82.033429}. They also provide a convenient framework to investigate non-Hermitian dynamics \Cite{Kawasaki2020PTSymmQW}. Here, we consider a QW of a single particle with an internal two-dimensional degree of freedom (the coin), spanned by $\lbrace{\ket{A},\ket{B} \rbrace}$. The evolution is discrete in time, with each step consisting of a coin rotation $W$, followed by a coin-dependent translation $T$:
\begin{equation}
U=TW, 
\label{eqn:singlestep}
\end{equation}
with $W=(\sigma_0+i\sigma_x)/\sqrt{2}$, where $\sigma_0$ and $\sigma_x$ are the identity and the first Pauli matrix, respectively, and
\begin{equation}
T=\alpha I+i\sum_m \left(\beta\ketbra{m+1,B}{m,A} +\text{h.c.}\right), 
\label{eqn:translation}
\end{equation}
where ${I=I_w\otimes \sigma_0}$ is the identity operator in the walker and coin space, $\alpha$ and $\beta$ are complex coefficients, and h.c.~stands for Hermitian conjugation. 
The lattice is spanned by position eigenstates $\lbrace\ket{m} \rbrace$, with $m$ an integer. The coupling coefficients between neighboring sites are reciprocal yet not complex conjugate to each other, which makes the evolution non-unitary. A visual representation of the action of the operator $T$ is provided in Fig. \figref{fig:fig1}(a).
Being translation invariant, the evolution can be conveniently diagonalized in the quasi-momentum space:
\begin{equation}
U=\int_\text{BZ}\text{d}q\,\mathcal{U}(q)\ketbra{q},
\label{eqn:bloch}
\end{equation}
where ${\ket{q}=\sum_m \exp(iqm)\ket{m}/\sqrt{2\pi}}$ is a quasi-momentum eigenstate, spanning the BZ ${[0,2\pi[}$. In Eq. \Eqref{eqn:bloch}, $\mathcal{U}$ can be modeled as generated by an effective Hamiltonian
\Cite{PhysRevA.82.033429},
\begin{equation}
\label{Hq}
    \mathcal{H}_{\rm eff}(q)=E(q)\textbf{n}(q)\cdot\bm{\sigma}\,,
\end{equation}
representing the central core of our analysis,
with
\begin{equation}
\mathcal{U}(q)=e^{-i\mathcal{H}_{\rm eff}(q)}\,.
\end{equation}
In Eq.\,\eqref{Hq}, $E$ is the quasi-energy band, $\textbf{n}$ is a complex-valued vector satisfying ${n_x^2+n_y^2+n_z^2=1}$ at each $q$, and $\bm{\sigma}=(\sigma_x,\sigma_y,\sigma_z)$ is the vector of the three Pauli matrices.
From Eqs. \Eqref{eqn:singlestep} and \Eqref{eqn:bloch}, we obtain
\begin{equation}
\begin{split}
E&=\arccos{\dfrac{\alpha-\beta\cos q}{\sqrt{2}}};\\
n_x&=-\dfrac{\alpha+\beta\cos q}{\sqrt{2-(\alpha-\beta\cos q)^2}};\\
n_y&=\dfrac{\beta\sin q}{\sqrt{2-(\alpha-\beta\cos q)^2}};\\
n_z&=\dfrac{\beta\sin q}{\sqrt{2-(\alpha-\beta\cos q)^2}}.
\end{split}
\end{equation}
We also define 
\begin{equation}
\begin{split}
\alpha&=\cos\frac{\delta+i\eta}{2};\\
\beta&=\sin\frac{\delta+i\eta}{2},
\label{eqn:mapping}
\end{split}
\end{equation}
where ${\delta \in [0,2\pi]}$ and $\eta$ encodes the degree of non-Hermiticity. Indeed, for ${\eta=0}$, the QW displays the conventional unitary evolution studied elsewhere \Cite{PhysRevResearch.2.023119}. 

The condition ${n_y=n_z}$ underlies sublattice symmetry \Cite{PhysRevX.9.041015}, i.e., there exists a unitary operator ${\mathcal{S}=\bm{s}\cdot\bm{\sigma}}$ such that ${\mathcal{S} \mathcal{H}(q) \mathcal{S}=-\mathcal{H}(q)}$. In our case, ${\bm{s}={(0,1,-1)}/\sqrt{2}}$. A topological invariant can thus be defined as the winding number \Cite{PhysRevA.97.052115,PhysRevX.9.041015}:
\begin{equation}
\nu=\dfrac{1}{2\pi}\int_\text{BZ}\text{d}q\,\left(\textbf{n}\times\dfrac{\partial \textbf{n}}{\partial q}\right)\cdot\bm{s}\,.
\label{eqn:winding}
\end{equation}
The topological diagram extracted from Eq. \Eqref{eqn:winding} for our QW protocol is plotted in Fig. \figref{fig:fig1}(b). 

{\it Experimental platform.---} The QW model described above is experimentally realized using the photonic platform introduced in Ref. \Cite{savarese2}. Here, walker states $\lbrace\ket{m}\rbrace$ are optical modes carrying $m$ units of transverse momentum: $\braket{x,y}{m}=A(x,y)\exp(im\Delta k)$, where $A(x,y)$ is the Gaussian beam envelope and ${\Delta k=2\pi/\Lambda}$ is a transverse-momentum unit along the $x$ direction, with $\Lambda$ a characteristic length \Cite{DErrico:20}. These modes can be resolved by recording the output light intensity in the focal plane of a lens, provided that ${w_0\geq\Lambda}$, with $w_0$ the beam waist \Cite{DErrico:20,savarese2}. Such optical encoding automatically maps the $x$ coordinate in real space into the walker quasi-momentum $q$, with the spatial period $\Lambda$ representing the first BZ. The coin states are mapped into left-handed ($\ket{\circlearrowleft}$) and right-handed ($\ket{\circlearrowright}$) circular polarization states. 

Both the coin rotation and the coin-dependent translation are realized via liquid-crystal metasurfaces (LCMSs). These are nematic samples with a patterned optic axis and electrically tunable birefringence \Cite{Rubano:19}. In the first case, a LCMS with uniform optic-axis orientation is employed, with its birefringence set at $\delta=\pi/2$. In this way, its action on light polarization corresponds to the operator $W$. The translation on the optical lattice is realized via a dichroic $g$-plate \Cite{savarese1,savarese2}. The latter features a periodic modulation of the optic axis, ${\theta(x)=\pi x/\Lambda}$, and is doped with a dichroic dye. The birefringence setting $\delta$ and the dichroic power $\eta$ of the $g$-plate are mapped into the QW hopping parameters via Eq. \Eqref{eqn:mapping}. By applying an AC voltage, the two parameters can be adjusted simultaneously, but not independently, as detailed in Ref. \Cite{savarese1}. The protocol for extracting $\delta$ and $\eta$ from polarimetric measurements at different voltages is reported in the Supplementary Material. 

\begin{figure}[t!]
    \centering
    \includegraphics[width=1.04\columnwidth]{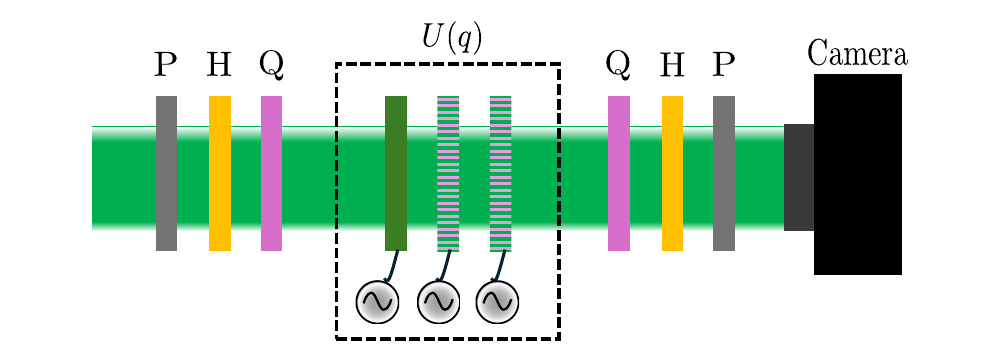}
    \caption{\textbf{Experimental process tomography.} Three liquid-crystal metasurfaces simulate a single QW step. The application of an external voltage allows for dynamically adjusting the parameters $\delta$ and $\eta$ of dichroic metasurfaces, encoding the model topology. Process tomography is realized by preparing and projecting onto the desired polarization states with a linear polarizer (P), a half-wave plate (H), and a quarter-wave plate (Q). Polarimetric images are then processed to retrieve the model eigenstructure.}
    \label{fig:fig2}
\end{figure}

\begin{figure*}[t!]
    \centering
    \includegraphics[width=0.88\linewidth]{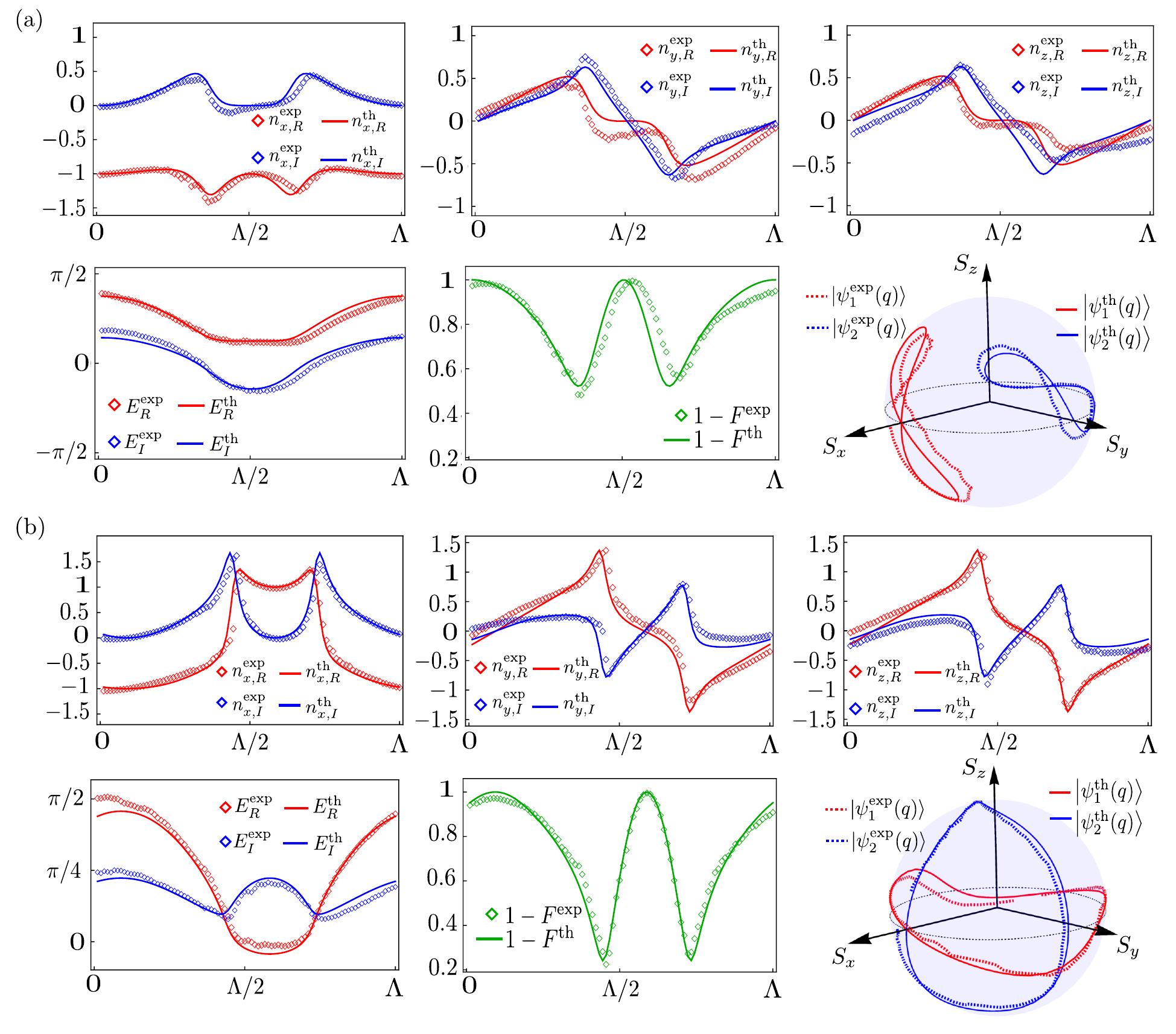}
    \caption{\textbf{Experimental results and theoretical predictions.} Tomographic reconstructions for the cases: (a)~${(\delta,\eta)=(\pi/4,0.9)}$, and (b)~${(\delta,\eta)=(1.3,1.4)}$ across one spatial period $\Lambda$, corresponding to the first BZ. Experimentally reconstructed real and imaginary parts of the energy bands and the $\textbf{n}$-vector components (diamonds) are compared with theoretical predictions (continuous lines). The overlap between the right eigenstates is also reported as infidelity at each quasi-momentum, and the trajectories of the polarization eigenstates are visualized on the Bloch sphere.}
    \label{fig:fig3}
\end{figure*}

To simulate $t$ steps of the QW, a sequence of $t$ stacks of single-step units is usually prepared, as demonstrated for instance in Ref. \Cite{savarese2}. In this paper, instead, we report the characterization of the eigenstructure of a single QW step across a full spatial period $\Lambda$, corresponding to a quasi-momentum-resolved process tomography over the BZ \Cite{dicolandrea2025}. The experimental setup is sketched in Fig. \figref{fig:fig2}. A 532-nm laser beam is expanded to cover approximately one BZ on the metasurfaces' plane, which corresponds to 5 mm in our setup. The reconstruction of the system eigenvalues and eigenstates is accomplished by numerically minimizing the distance between a set of experimental and theoretical polarimetric measurements, expressed as functions of the quasi-energy $E$ and the $\textbf{n}$-vector components. In the following, we will always refer to the right eigenstates of the non-Hermitian Hamiltonian. Each polarimetric measurement is realized by setting an input polarization state $\ket{i}$, letting it evolve under $U$, and recording the light intensity after projecting onto $\ket{j}$:
\begin{equation}
I_\text{ij}=I_0\abs{\bra{j}U\ket{i}}^2, 
\label{eqn:polarimetry}
\end{equation}
where $I_0$ is the total intensity of the light beam. A linear polarizer (P), a half-wave plate (HWP), and a quarter-wave plate (QWP) are used to prepare the desired input polarization state. The beam propagates through the LCMSs simulating the QW. To enhance the dichroic character of a single step and probe a broader region of the topological diagram (see Fig. \figref{fig:fig1}(b)), we cascade two $g$-plates, tuned at $\delta_1$ and $\delta_2$, respectively, so that their total birefringence amounts to the target setting: ${\delta=\delta_1+\delta_2}$. The tomographic measurement is completed by projecting the output beam onto a specific polarization state with the reverse sequence QWP-HWP-P. Light intensity distribution is recorded on a camera placed after the projection stage to minimize propagation effects. Polarimetric images are first compressed into ${\left(90\times90\right)}$-pixel grids to average over local intensity fluctuations and then processed to extract the system parameters. A further noise-reducing average is obtained by integrating the light intensity along the $y$ direction. A detailed description of the experimental setup, along with the complete set of polarimetric measurements and technical details on our optimization routine, can be found in the Supplementary Material. 

{\it Results.---} Our photonic encoding enables direct probing of the system in reciprocal space. Consequently, the process tomography is performed by directly scanning the $q$-space across one BZ, with each pixel cluster corresponding to a discretized quasi-momentum value. We stress that the inherently easy access to the BZ granted by our platform would normally require further adjustments to the setup in other standard photonic encodings, such as QWs performed in real space with beam displacers \Cite{Xiao2017,PhysRevLett.119.130501,Wang2019,Xiao2020,PhysRevLett.126.230402,PhysRevLett.133.070801,Zhang2025} or in the time domain within fiber loops \Cite{Weidemann2020,Weidemann2022,Lin2022,PhysRevLett.132.203801,Xue2024}, with losses realized through partial measurements on the photon polarization. 

We provide tomographic reconstructions for different settings of the ${\delta-\eta}$ parameters, corresponding to topologically trivial or non-trivial regimes. In the following, $\delta$ and $\eta$ refer to the sum of individual birefringence and dichroism settings of the two $g$-plates employed in the experiment. Figure \figref{fig:fig3} shows the real (red curves) and imaginary (blue curves) parts of the reconstructed quasi-energy $E$ and the $\textbf{n}$-vector components for the cases (a) $(\delta,\eta)=(\pi/4,0.9)$ and (b) $(\delta,\eta)=(1.3,1.4)$. The experimental reconstructions (empty diamonds) are compared with the theoretical predictions (solid lines). At each quasi-momentum value, the agreement with the theory is quantified by the operator fidelity
\begin{equation}
\mathcal{F}=\abs{\dfrac{\text{Tr}\left(U^\dagger_\text{th}U_\text{exp} \right)}{\sqrt{U^\dagger_\text{th}U_\text{th}}\sqrt{U^\dagger_\text{exp}U_\text{exp}}}},
\end{equation}
where $U_\text{th}$ and $U_\text{exp}$ are the theoretical and experimentally reconstructed processes. The average fidelities for the cases considered in Fig. \figref{fig:fig3} are $(98.8\pm0.5)\%$ and $(99\pm2)\%$, respectively, where the average is taken over all pixels and the error is estimated as one standard deviation.

By diagonalizing the reconstructed Hamiltonian at each $q$, we extract the right eigenstates, ${\ket{\psi_1(q)}}$ and ${\ket{\psi_2(q)}}$, for which we compute the state fidelity: ${F=\abs{\braket{\psi_{1}}{\psi_{2}}}^2}$. In the form of infidelity ${1-F}$, the latter is reported in Fig. \figref{fig:fig3} as green diamonds (experiment) and solid lines (theory). By mapping the Hamiltonian eigenstates to 3D vectors via ${S_{i,{1;2}}=\mel{\sigma_i}{\psi_{1;2}}{\sigma_i}}$, where ${\textbf{S}=(S_x,S_y,S_z)}$ defines the Stokes parameters of the polarization eigenstates, we can follow the trajectories on the Bloch-Poincaré sphere, drawn as $q$ runs across the BZ. The non-perfect periodic character of the experimentally reconstructed eigenstates must be ascribed to alignment imperfections and fabrication defects.  

Increasing $\eta$ in our system corresponds to enhancing the non-unitary character of the process. This is observed in Fig. \figref{fig:fig3}(a), where all the features exhibit non-negligible imaginary contributions, and the infidelity shows two minima, corresponding to the \emph{exceptional points}, which degenerate into full eigenstate coalescence across the topological phase transition line. At ${\delta=1.3}$, the topological phase transition occurs at $\eta\simeq0.99$. For a fixed $\delta$, the dichroic parameter can thus be tuned to also drive the system across different topological phases, switching from trivial to non-trivial regimes, as illustrated in Fig. \figref{fig:fig1}(b). This scenario is realized in Fig. \figref{fig:fig3}(b), which shows large imaginary contributions and cusps at the exceptional points. An experimental realization for a low value of $\eta$, corresponding to a weakly non-Hermitian regime, is presented in Appendix A. To assess the system topology, we numerically evaluate the integral of Eq. \Eqref{eqn:winding}. From the reconstructed $\textbf{n}$ vectors, we obtain $0.03$ and $0.95$ for the cases (a) and (b), respectively, in agreement with theoretical predictions (cf.~Fig. \figref{fig:fig1}(b)). Topologically trivial bands are associated with self-intersecting trajectories on the Bloch sphere (see Fig. \figref{fig:fig3}(a)), while closed, non-overlapping loops are observed for the non-trivial case of Fig. \figref{fig:fig3}(b). 

\begin{figure}[t!]
\centering
\includegraphics[width=.88\columnwidth]{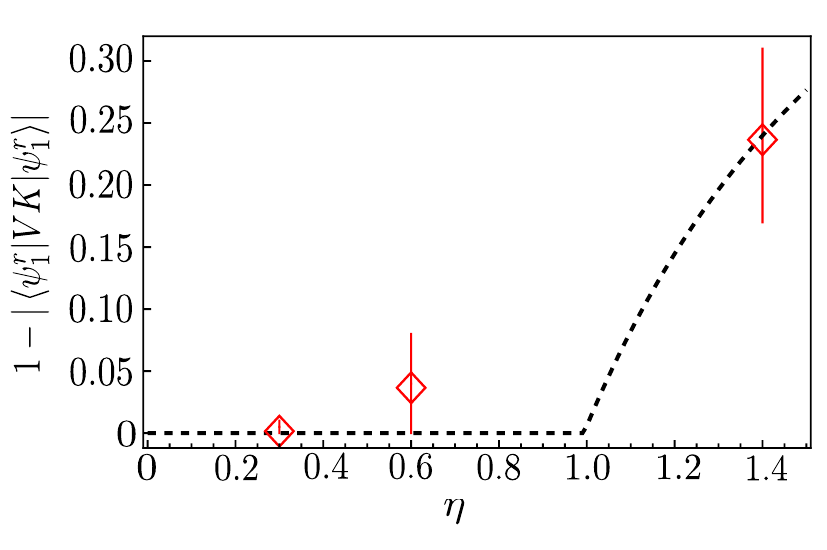}
\caption{\textbf{PT-symmetry breaking.} 
The system is driven through a topological phase transition by tuning $\eta$ from 0.3 to 0.6 to 1.3, keeping ${\delta=1.3}$. At the critical value of quasi-momentum $q_c$, PT-symmetry breaking is revealed by the order parameter ${1 - \big|\langle \psi_{1}^r | VK | \psi_{1}^r \rangle\big|}$, whose theoretical prediction is drawn as dashed line. The red diamonds are obtained from the experimentally reconstructed Hamiltonian eigenstates (see also Fig.\,\ref{fig:pt-appendix} in Appendix B).
}
    \label{fig:fig4}
\end{figure}

Finally, we reveal the breaking of the PT symmetry across a phase transition. We start by rotating the Hamiltonian ${\mathcal{H}_r=r\mathcal{H}r^\dagger}$ so as ${\mathcal{S}=\sigma_z}$, which implies $r=\sqrt{W}$.

\color{black}

At the critical value of the quasi-momentum $q_c$, the rotated Hamiltonian reads
\begin{equation}\label{rotHam}
H_r=\begin{pmatrix}
0 & ae^{i\phi}\\
be^{-i\phi} & 0
\end{pmatrix},
\end{equation}
where $a$, $b$, and $\phi$ are real numbers. We then evaluate the order parameter $1-\abs{\mel{\psi_{1;2}^r}{VK}{\psi_{1;2}^r}}$, where ${V=\cos{\phi}\sigma_0+i\sin{\phi}\sigma_z}$ and $K$ is the complex conjugation operator, on the eigenstates of the rotated Hamiltonian. Here, the operator $VK$ plays the role of a rotated PT symmetry, being unitarily equivalent to the PT symmetry operator (see Appendix B for the complete derivation). Despite $VK$ commutes with the effective Hamiltonian of Eq. \Eqref{rotHam}, the eigenstates break the symmetry for sufficiently high values of $\eta$, corresponding to topologically non-trivial phases. We experimentally drive the system through a topological phase transition by keeping ${\delta=1.3}$ and tuning $\eta$ from $0.3$ to $0.6$ to $1.4$ (the latter setting has already been investigated in Fig. \figref{fig:fig3}(b)). Experimental data obtained for the three cases are plotted in Fig. \figref{fig:fig4}, where each datapoint is obtained as the average of the three pixel clusters corresponding to the minimum fidelity, and the error bar is given as the standard deviation. Good agreement is observed in the unbroken phase (${\eta\lesssim 0.99}$), with larger errors appearing in correspondence with the symmetry breaking at ${\eta\simeq1.4}$. The complete tomographic reconstructions for the cases $(\delta,\eta)=(1.3,0.3)$ and $(\delta,\eta)=(1.3,0.6)$ can be found in the Supplementary Material.  
We also note that the Hamiltonian in the form of Eq. \Eqref{rotHam} necessarily admits two opposite real eigenvalues in the case ${ab>0}$ and purely imaginary if ${ab<0}$. The first case occurs for ${\eta<\eta_c}$ corresponding to the topologically trivial phase and the unbroken $VK$-phase. Approaching the topological phase and the broken $VK$-phase, we find opposite and purely imaginary eigenvalues, corresponding to a balanced gain-loss mechanism. The sudden crossing between a phase with opposite real eigenvalues and a phase with opposite purely imaginary eigenvalues represents a direct consequence of the simultaneous presence of $VK$-symmetry (rotated PT symmetry) and sublattice symmetry (see Fig.\,\ref{fig:pt-appendix} in Appendix~B for details).
\color{black}

 

{\it Discussion.---} In conclusion, we have shown that non-Hermitian topological Hamiltonians can be engineered and probed experimentally by exploiting reciprocal-space resolution in a compact photonic quantum-walk platform. We have implemented an optical simulator of a single step and performed a quasi-momentum-resolved tomographic characterization across one Brillouin Zone.
The accuracy of the reconstruction is inferred by the comparison between experimental measurements and theoretical expectations. 
Remarkably, our platform provides tunable parameters that encode the hopping and loss coefficients, allowing us to dynamically control the system topology. 

The tomography enabled the direct mapping of non-Hermitian band structures, the detection of exceptional points in momentum space, and the observation of PT-symmetry breaking through eigenstate coalescence. From the reconstructed eigenstates, the topological invariant of the system has also been determined. 

Future experimental studies will concern the evolution of dynamical observables to track the winding number, as already demonstrated for unitary walks \Cite{Cardano2017}. The increasing experimental complexity might be tackled by adapting the approach of Ref. \Cite{DiColandreaQW} to the non-unitary simulation regime. A further stimulating avenue will be the engineering of 
Liouvillian exceptional points in this setting and the exploration of 
thermodynamic features across PT-symmetry phase transitions.

\bigskip
{\it Acknowledgments.---}
This work was supported by the PNRR MUR project PE0000023-NQSTI.

{\it Disclosures.---}
The authors declare no conflicts of interest.

\clearpage

\bibliography{apssamp, bib-notes}

@article{ashida2020non,
  title={Non-hermitian physics},
  author={Ashida, Yuto and Gong, Zongping and Ueda, Masahito},
  journal={Adv. Phys.},
  volume={69},
  number={3},
  pages={249--435},
  year={2020},
  publisher={Taylor \& Francis},
url = {https://www.tandfonline.com/doi/abs/10.1080/00018732.2021.1876991}
}

@article{PhysRevA.100.062131,
  title = {Quantum exceptional points of non-Hermitian Hamiltonians and Liouvillians: The effects of quantum jumps},
  author = {Minganti, Fabrizio and Miranowicz, Adam and Chhajlany, Ravindra W. and Nori, Franco},
  journal = {Phys. Rev. A},
  volume = {100},
  issue = {6},
  pages = {062131},
  numpages = {17},
  year = {2019},
  month = {Dec},
  publisher = {American Physical Society},
  doi = {10.1103/PhysRevA.100.062131},
  url = {https://link.aps.org/doi/10.1103/PhysRevA.100.062131}
}

@article{savarese1,
    author = {Savarese, Paola and Bansal, Sarvesh and Ammendola, Maria Gorizia and Barboza, Raouf and Salvatore, Marcella and Oscurato, Stefano Luigi and Piccirillo, Bruno and Di Colandrea, Francesco and Marrucci, Lorenzo and Cardano, Filippo},
    title = {Electrically tunable liquid-crystal metasurfaces with patterned birefringence and dichroism},
    journal = {APL Photonics},
    volume = {10},
    number = {5},
    pages = {050802},
    year = {2025},
    month = {05},
    issn = {2378-0967},
    doi = {10.1063/5.0261491},
    url = {https://doi.org/10.1063/5.0261491}
}

@article{savarese2,
    author = {Savarese, Paola and Bansal, Sarvesh and Ammendola, Maria Gorizia and Amato, Lorenzo and Barboza, Raouf and Piccirillo, Bruno and Di Colandrea, Francesco and Marrucci, Lorenzo and Cardano, Filippo},
    title = {Programmable non-Hermitian photonic quantum walks via dichroic metasurfaces},
    journal = {APL Photonics},
    volume = {10},
    number = {8},
    pages = {086106},
    year = {2025},
    month = {08},
    issn = {2378-0967},
    doi = {10.1063/5.0274332},
    url = {https://doi.org/10.1063/5.0274332}
}

@article{PhysRevX.9.041015,
  title = {Symmetry and Topology in Non-Hermitian Physics},
  author = {Kawabata, Kohei and Shiozaki, Ken and Ueda, Masahito and Sato, Masatoshi},
  journal = {Phys. Rev. X},
  volume = {9},
  issue = {4},
  pages = {041015},
  numpages = {52},
  year = {2019},
  month = {Oct},
  publisher = {American Physical Society},
  doi = {10.1103/PhysRevX.9.041015},
  url = {https://link.aps.org/doi/10.1103/PhysRevX.9.041015}
}

@article{PhysRevLett.133.180801,
  title = {Non-Hermitian Sensing in the Absence of Exceptional Points},
  author = {Xiao, Lei and Chu, Yaoming and Lin, Quan and Lin, Haiqing and Yi, Wei and Cai, Jianming and Xue, Peng},
  journal = {Phys. Rev. Lett.},
  volume = {133},
  issue = {18},
  pages = {180801},
  numpages = {7},
  year = {2024},
  month = {Oct},
  publisher = {American Physical Society},
  doi = {10.1103/PhysRevLett.133.180801},
  url = {https://link.aps.org/doi/10.1103/PhysRevLett.133.180801}
}

@article{naghiloo2019quantum,
  title={Quantum state tomography across the exceptional point in a single dissipative qubit},
  author={Naghiloo, M and Abbasi, M and Joglekar, Yogesh N and Murch, KW},
  journal={Nat. Phys.},
  volume={15},
  number={12},
  pages={1232--1236},
  year={2019},
  publisher={Nature Publishing Group UK London},
url = {https://www.nature.com/articles/s41567-019-0652-z}
}

@article{DErrico:20,
author = {Alessio D'Errico and Filippo Cardano and Maria Maffei and Alexandre Dauphin and Raouf Barboza and Chiara Esposito and Bruno Piccirillo and Maciej Lewenstein and Pietro Massignan and Lorenzo Marrucci},
journal = {Optica},
number = {2},
pages = {108--114},
publisher = {Optica Publishing Group},
title = {Two-dimensional topological quantum walks in the momentum space of structured light},
volume = {7},
month = {Feb},
year = {2020},
url = {https://opg.optica.org/optica/abstract.cfm?URI=optica-7-2-108},
doi = {10.1364/OPTICA.365028},
}

@article{Weidemann2022,
  author    = {Sebastian Weidemann and Mark Kremer and Stefano Longhi and Alexander Szameit},
  title     = {Topological triple phase transition in non-Hermitian Floquet quasicrystals},
  journal   = {Nature},
  year      = {2022},
  volume    = {601},
  number    = {7893},
  pages     = {354--359},
  doi       = {10.1038/s41586-021-04253-0},
  url       = {https://doi.org/10.1038/s41586-021-04253-0},
  issn      = {1476-4687}
}

@article{Lin2022,
  author    = {Quan Lin and Tianyu Li and Lei Xiao and Kunkun Wang and Wei Yi and Peng Xue},
  title     = {Observation of non-Hermitian topological Anderson insulator in quantum dynamics},
  journal   = {Nat. Commun.},
  year      = {2022},
  volume    = {13},
  number    = {1},
  pages     = {3229},
  doi       = {10.1038/s41467-022-30938-9},
  url       = {https://doi.org/10.1038/s41467-022-30938-9},
  issn      = {2041-1723}
}

@article{DiColandreaQW,
author = {Francesco {Di Colandrea} and Amin Babazadeh and Alexandre Dauphin and Pietro Massignan and Lorenzo Marrucci and Filippo Cardano},
journal = {Optica},
number = {3},
pages = {324--331},
publisher = {Optica Publishing Group},
title = {Ultra-long quantum walks via spin--orbit photonics},
volume = {10},
month = {Mar},
year = {2023},
url = {https://opg.optica.org/optica/abstract.cfm?URI=optica-10-3-324},
doi = {10.1364/OPTICA.474542}
}

@article{Cardano2017,
  author    = {Filippo Cardano and Alessio D’Errico and Alexandre Dauphin and Maria Maffei and Bruno Piccirillo and Corrado de Lisio and Giulio De Filippis and Vittorio Cataudella and Enrico Santamato and Lorenzo Marrucci and Maciej Lewenstein and Pietro Massignan},
  title     = {Detection of Zak phases and topological invariants in a chiral quantum walk of twisted photons},
  journal   = {Nat. Commun.},
  year      = {2017},
  volume    = {8},
  number    = {1},
  pages     = {15516},
  doi       = {10.1038/ncomms15516},
  url       = {https://doi.org/10.1038/ncomms15516},
  issn      = {2041-1723}
}

@article{Piccirillo2010,
  author  = {Piccirillo, Bruno and D’Ambrosio, Vincenzo and Slussarenko, Sergei and Marrucci, Lorenzo and Santamato, Enrico},
  title   = {Photon spin-to-orbital angular momentum conversion via an electrically tunable q-plate},
  journal = {Appl. Phys. Lett.},
  volume  = {97},
  number  = {24},
  pages   = {241104},
  year    = {2010},
  month   = dec,
  doi     = {10.1063/1.3527083},
  url     = {https://doi.org/10.1063/1.3527083}
}

@article{
Weidemann2020,
author = {Sebastian Weidemann  and Mark Kremer  and Tobias Helbig  and Tobias Hofmann  and Alexander Stegmaier  and Martin Greiter  and Ronny Thomale  and Alexander Szameit },
title = {Topological funneling of light},
journal = {Science},
volume = {368},
number = {6488},
pages = {311-314},
year = {2020},
doi = {10.1126/science.aaz8727},
URL = {https://www.science.org/doi/abs/10.1126/science.aaz8727}}

@article{Kawasaki2020PTSymmQW,
  author  = {Kawasaki, Makio and Mochizuki, Ken and Kawakami, Norio and Obuse, Hideaki},
  title   = {Bulk--edge correspondence and stability of multiple edge states of a $\mathcal{PT}$-symmetric non-Hermitian system by using non-unitary quantum walks},
  journal = {Prog. Theor. Exp. Phys.},
  volume  = {2020},
  number  = {12},
  pages   = {12A105},
  year    = {2020},
  month   = {07},
  issn    = {2050-3911},
  doi     = {10.1093/ptep/ptaa034},
  url     = {https://doi.org/10.1093/ptep/ptaa034}
}

@article{PhysRevResearch.2.023119,
  title = {Bulk detection of time-dependent topological transitions in quenched chiral models},
  author = {D'Errico, Alessio and Di Colandrea, Francesco and Barboza, Raouf and Dauphin, Alexandre and Lewenstein, Maciej and Massignan, Pietro and Marrucci, Lorenzo and Cardano, Filippo},
  journal = {Phys. Rev. Res.},
  volume = {2},
  issue = {2},
  pages = {023119},
  numpages = {6},
  year = {2020},
  month = {May},
  publisher = {American Physical Society},
  doi = {10.1103/PhysRevResearch.2.023119},
  url = {https://link.aps.org/doi/10.1103/PhysRevResearch.2.023119}
}

@article{MARES2020126302,
title = {Quantum walk transport on carbon nanotube structures},
journal = {Phys. Lett. A},
volume = {384},
number = {15},
pages = {126302},
year = {2020},
issn = {0375-9601},
doi = {https://doi.org/10.1016/j.physleta.2020.126302},
url = {https://www.sciencedirect.com/science/article/pii/S0375960120301018},
author = {J. Mareš and J. Novotný and I. Jex}
}

@article{PhysRevA.82.033429,
  title = {Exploring topological phases with quantum walks},
  author = {Kitagawa, Takuya and Rudner, Mark S. and Berg, Erez and Demler, Eugene},
  journal = {Phys. Rev. A},
  volume = {82},
  issue = {3},
  pages = {033429},
  numpages = {9},
  year = {2010},
  month = {Sep},
  publisher = {American Physical Society},
  doi = {10.1103/PhysRevA.82.033429},
  url = {https://link.aps.org/doi/10.1103/PhysRevA.82.033429}
}

@article{PhysRevLett.132.203801,
  title = {Observation of Non-Hermitian Edge Burst Effect in One-Dimensional Photonic Quantum Walk},
  author = {Zhu, Jiankun and Mao, Ya-Li and Chen, Hu and Yang, Kui-Xing and Li, Linhu and Yang, Bing and Li, Zheng-Da and Fan, Jingyun},
  journal = {Phys. Rev. Lett.},
  volume = {132},
  issue = {20},
  pages = {203801},
  numpages = {6},
  year = {2024},
  month = {May},
  publisher = {American Physical Society},
  doi = {10.1103/PhysRevLett.132.203801},
  url = {https://link.aps.org/doi/10.1103/PhysRevLett.132.203801}
}

@article{PhysRevLett.133.070801,
  title = {Observation of Non-Hermitian Edge Burst in Quantum Dynamics},
  author = {Xiao, Lei and Xue, Wen-Tan and Song, Fei and Hu, Yu-Min and Yi, Wei and Wang, Zhong and Xue, Peng},
  journal = {Phys. Rev. Lett.},
  volume = {133},
  issue = {7},
  pages = {070801},
  numpages = {6},
  year = {2024},
  month = {Aug},
  publisher = {American Physical Society},
  doi = {10.1103/PhysRevLett.133.070801},
  url = {https://link.aps.org/doi/10.1103/PhysRevLett.133.070801}
}

@article{Xue2024,
  author    = {Peng Xue and Quan Lin and Kunkun Wang and Lei Xiao and Stefano Longhi and Wei Yi},
  title     = {Self acceleration from spectral geometry in dissipative quantum-walk dynamics},
  journal   = {Nat. Commun.},
  year      = {2024},
  volume    = {15},
  number    = {1},
  pages     = {4381},
  doi       = {10.1038/s41467-024-48815-y},
  url       = {https://doi.org/10.1038/s41467-024-48815-y},
  issn      = {2041-1723}
}

@article{Zhang2025,
  author    = {Haiting Zhang and Kunkun Wang and Lei Xiao and Peng Xue},
  title     = {Self-normal and biorthogonal dynamical quantum phase transitions in non-Hermitian quantum walks},
  journal   = {Light: Sci. Appl.},
  year      = {2025},
  volume    = {14},
  number    = {1},
  pages     = {253},
  doi       = {10.1038/s41377-025-01919-6},
  url       = {https://doi.org/10.1038/s41377-025-01919-6},
  issn      = {2047-7538}
}

@article{Wang2019,
  author    = {Kunkun Wang and Xingze Qiu and Lei Xiao and Xiang Zhan and Zhihao Bian and Barry C. Sanders and Wei Yi and Peng Xue},
  title     = {Observation of emergent momentum--time skyrmions in parity--time-symmetric non-unitary quench dynamics},
  journal   = {Nat. Commun.},
  year      = {2019},
  volume    = {10},
  number    = {1},
  pages     = {2293},
  doi       = {10.1038/s41467-019-10252-7},
  url       = {https://doi.org/10.1038/s41467-019-10252-7},
  issn      = {2041-1723}
}

@article{PhysRevLett.126.230402,
  title = {Observation of Non-Bloch Parity-Time Symmetry and Exceptional Points},
  author = {Xiao, Lei and Deng, Tianshu and Wang, Kunkun and Wang, Zhong and Yi, Wei and Xue, Peng},
  journal = {Phys. Rev. Lett.},
  volume = {126},
  issue = {23},
  pages = {230402},
  numpages = {6},
  year = {2021},
  month = {Jun},
  publisher = {American Physical Society},
  doi = {10.1103/PhysRevLett.126.230402},
  url = {https://link.aps.org/doi/10.1103/PhysRevLett.126.230402}
}

@article{PhysRevResearch.6.023202,
  title = {Quantum metric of non-Hermitian Su-Schrieffer-Heeger systems},
  author = {Chen Ye, Chao and Vleeshouwers, W. L. and Heatley, S. and Gritsev, V. and Morais Smith, C.},
  journal = {Phys. Rev. Res.},
  volume = {6},
  issue = {2},
  pages = {023202},
  numpages = {14},
  year = {2024},
  month = {May},
  publisher = {American Physical Society},
  doi = {10.1103/PhysRevResearch.6.023202},
  url = {https://link.aps.org/doi/10.1103/PhysRevResearch.6.023202}
}

@article{
cardanoscience,
author = {Filippo Cardano  and Francesco Massa  and Hammam Qassim  and Ebrahim Karimi  and Sergei Slussarenko  and Domenico Paparo  and Corrado de Lisio  and Fabio Sciarrino  and Enrico Santamato  and Robert W. Boyd  and Lorenzo Marrucci },
title = {Quantum walks and wavepacket dynamics on a lattice with twisted photons},
journal = {Sci. Adv.},
volume = {1},
number = {2},
pages = {e1500087},
year = {2015},
doi = {10.1126/sciadv.1500087},
URL = {https://www.science.org/doi/abs/10.1126/sciadv.1500087}}

@article{Venegas_Andraca_2012,
   title={Quantum walks: a comprehensive review},
   volume={11},
   ISSN={1573-1332},
   url={http://dx.doi.org/10.1007/s11128-012-0432-5},
   DOI={10.1007/s11128-012-0432-5},
   number={5},
   journal={Quantum Inf. Process.},
   publisher={Springer Science and Business Media LLC},
   author={Venegas-Andraca, Salvador Elías},
   year={2012},
   month=jul, pages={1015–1106} }

@article{Xiao2017,
  author    = {L. Xiao and X. Zhan and Z. H. Bian and K. K. Wang and X. Zhang and X. P. Wang and J. Li and K. Mochizuki and D. Kim and N. Kawakami and W. Yi and H. Obuse and B. C. Sanders and P. Xue},
  title     = {Observation of topological edge states in parity--time-symmetric quantum walks},
  journal   = {Nat. Phys.},
  year      = {2017},
  volume    = {13},
  number    = {11},
  pages     = {1117--1123},
  doi       = {10.1038/nphys4204},
  url       = {https://doi.org/10.1038/nphys4204},
  issn      = {1745-2481}
}

@article{Xiao2020,
  author    = {Lei Xiao and Tianshu Deng and Kunkun Wang and Gaoyan Zhu and Zhong Wang and Wei Yi and Peng Xue},
  title     = {Non-Hermitian bulk--boundary correspondence in quantum dynamics},
  journal   = {Nat. Phys.},
  year      = {2020},
  volume    = {16},
  number    = {7},
  pages     = {761--766},
  doi       = {10.1038/s41567-020-0836-6},
  url       = {https://doi.org/10.1038/s41567-020-0836-6},
  issn      = {1745-2481}
}

@article{PhysRevLett.119.130501,
  title = {Detecting Topological Invariants in Nonunitary Discrete-Time Quantum Walks},
  author = {Zhan, Xiang and Xiao, Lei and Bian, Zhihao and Wang, Kunkun and Qiu, Xingze and Sanders, Barry C. and Yi, Wei and Xue, Peng},
  journal = {Phys. Rev. Lett.},
  volume = {119},
  issue = {13},
  pages = {130501},
  numpages = {6},
  year = {2017},
  month = {Sep},
  publisher = {American Physical Society},
  doi = {10.1103/PhysRevLett.119.130501},
  url = {https://link.aps.org/doi/10.1103/PhysRevLett.119.130501}
}

@misc{dicolandrea2025,
      title={Skyrmionic polarization textures in structured dielectric planar media}, 
      author={Francesco {Di Colandrea} and Lorenzo Marrucci and Filippo Cardano},
      archivePrefix={arXiv},
      url={https://arxiv.org/abs/2510.09427},
journal = {arxiv.org/abs/2510.09427}
}

@article{Rubano:19,
author = {Andrea Rubano and Filippo Cardano and Bruno Piccirillo and Lorenzo Marrucci},
journal = {J. Opt. Soc. Am. B},
number = {5},
pages = {D70--D87},
publisher = {Optica Publishing Group},
title = {Q-plate technology: a progress review [Invited]},
volume = {36},
month = {May},
year = {2019},
url = {https://opg.optica.org/josab/abstract.cfm?URI=josab-36-5-D70},
doi = {10.1364/JOSAB.36.000D70},
}

@article{PhysRevA.97.052115,
  title = {Geometrical meaning of winding number and its characterization of topological phases in one-dimensional chiral non-Hermitian systems},
  author = {Yin, Chuanhao and Jiang, Hui and Li, Linhu and L\"u, Rong and Chen, Shu},
  journal = {Phys. Rev. A},
  volume = {97},
  issue = {5},
  pages = {052115},
  numpages = {7},
  year = {2018},
  month = {May},
  publisher = {American Physical Society},
  doi = {10.1103/PhysRevA.97.052115},
  url = {https://link.aps.org/doi/10.1103/PhysRevA.97.052115}
}

@article{RevModPhys.96.045002,
  title = {$\mathcal{PT}$-symmetric quantum mechanics},
  author = {Bender, Carl M. and Hook, Daniel W.},
  journal = {Rev. Mod. Phys.},
  volume = {96},
  issue = {4},
  pages = {045002},
  numpages = {54},
  year = {2024},
  month = {Oct},
  publisher = {American Physical Society},
  doi = {10.1103/RevModPhys.96.045002},
  url = {https://link.aps.org/doi/10.1103/RevModPhys.96.045002}
}

@article{lu2025dynamical,
  title={Dynamical topology of chiral and nonreciprocal state transfers in a non-Hermitian quantum system},
  author={Lu, Pengfei and Liu, Yang and Lao, Qifeng and Liu, Teng and Rao, Xinxin and Bian, Ji and Wu, Hao and Zhu, Feng and Luo, Le},
  journal={Commun. Phys.},
  volume={8},
  number={1},
  pages={91},
  year={2025},
  publisher={Nature Publishing Group UK London},
url = {https://www.nature.com/articles/s42005-025-01989-3}
}

@article{roccati2022non,
  title={Non-Hermitian physics and master equations},
  author={Roccati, Federico and Palma, G Massimo and Ciccarello, Francesco and Bagarello, Fabio},
  journal={Open Syst. Inf. Dyn.},
  volume={29},
  number={01},
  pages={2250004},
  year={2022},
  publisher={World Scientific},
url = {https://www.worldscientific.com/doi/10.1142/S1230161222500044}
}

@article{PhysRevLett.123.170401,
  title = {Non-Hermitian Skin Effect and Chiral Damping in Open Quantum Systems},
  author = {Song, Fei and Yao, Shunyu and Wang, Zhong},
  journal = {Phys. Rev. Lett.},
  volume = {123},
  issue = {17},
  pages = {170401},
  numpages = {8},
  year = {2019},
  month = {Oct},
  publisher = {American Physical Society},
  doi = {10.1103/PhysRevLett.123.170401},
  url = {https://link.aps.org/doi/10.1103/PhysRevLett.123.170401}
}

@article{PhysRevB.95.201407,
  title = {Detecting topological invariants in chiral symmetric insulators via losses},
  author = {Rakovszky, Tibor and Asb\'oth, J\'anos K. and Alberti, Andrea},
  journal = {Phys. Rev. B},
  volume = {95},
  issue = {20},
  pages = {201407},
  numpages = {6},
  year = {2017},
  month = {May},
  publisher = {American Physical Society},
  doi = {10.1103/PhysRevB.95.201407},
  url = {https://link.aps.org/doi/10.1103/PhysRevB.95.201407}
}

@article{doppler2016dynamically,
  title={Dynamically encircling an exceptional point for asymmetric mode switching},
  author={Doppler, J{\"o}rg and Mailybaev, Alexei A and B{\"o}hm, Julian and Kuhl, Ulrich and Girschik, Adrian and Libisch, Florian and Milburn, Thomas J and Rabl, Peter and Moiseyev, Nimrod and Rotter, Stefan},
  journal={Nature},
  volume={537},
  number={7618},
  pages={76--79},
  year={2016},
  publisher={Nature Publishing Group UK London},
url = {https://www.nature.com/articles/nature18605}
}

@article{PhysRevLett.102.065703,
  title = {Topological Transition in a Non-Hermitian Quantum Walk},
  author = {Rudner, M. S. and Levitov, L. S.},
  journal = {Phys. Rev. Lett.},
  volume = {102},
  issue = {6},
  pages = {065703},
  numpages = {4},
  year = {2009},
  month = {Feb},
  publisher = {American Physical Society},
  doi = {10.1103/PhysRevLett.102.065703},
  url = {https://link.aps.org/doi/10.1103/PhysRevLett.102.065703}
}

@article{el2018non,
  title={Non-Hermitian physics and PT symmetry},
  author={El-Ganainy, Ramy and Makris, Konstantinos G and Khajavikhan, Mercedeh and Musslimani, Ziad H and Rotter, Stefan and Christodoulides, Demetrios N},
  journal={Nat. Phys.},
  volume={14},
  number={1},
  pages={11--19},
  year={2018},
  publisher={Nature Publishing Group UK London},
url = {https://www.nature.com/articles/nphys4323}
}

@article{PhysRevLett.103.093902,
  title = {Observation of $\mathcal{P}\mathcal{T}$-Symmetry Breaking in Complex Optical Potentials},
  author = {Guo, A. and Salamo, G. J. and Duchesne, D. and Morandotti, R. and Volatier-Ravat, M. and Aimez, V. and Siviloglou, G. A. and Christodoulides, D. N.},
  journal = {Phys. Rev. Lett.},
  volume = {103},
  issue = {9},
  pages = {093902},
  numpages = {4},
  year = {2009},
  month = {Aug},
  publisher = {American Physical Society},
  doi = {10.1103/PhysRevLett.103.093902},
  url = {https://link.aps.org/doi/10.1103/PhysRevLett.103.093902}
}

@article{PhysRevX.8.031079,
  title = {Topological Phases of Non-Hermitian Systems},
  author = {Gong, Zongping and Ashida, Yuto and Kawabata, Kohei and Takasan, Kazuaki and Higashikawa, Sho and Ueda, Masahito},
  journal = {Phys. Rev. X},
  volume = {8},
  issue = {3},
  pages = {031079},
  numpages = {33},
  year = {2018},
  month = {Sep},
  publisher = {American Physical Society},
  doi = {10.1103/PhysRevX.8.031079},
  url = {https://link.aps.org/doi/10.1103/PhysRevX.8.031079}
}

@article{PhysRevLett.132.156901,
  title = {Probing $PT$-Symmetry Breaking of Non-Hermitian Topological Photonic States via Strong Photon-Magnon Coupling},
  author = {Qian, Jie and Li, Jie and Zhu, Shi-Yao and You, J. Q. and Wang, Yi-Pu},
  journal = {Phys. Rev. Lett.},
  volume = {132},
  issue = {15},
  pages = {156901},
  numpages = {6},
  year = {2024},
  month = {Apr},
  publisher = {American Physical Society},
  doi = {10.1103/PhysRevLett.132.156901},
  url = {https://link.aps.org/doi/10.1103/PhysRevLett.132.156901}
}

@article{bender1998real,
  title = {Real Spectra in Non-Hermitian Hamiltonians Having ${PT}$ Symmetry},
  author = {Bender, Carl M. and Boettcher, Stefan},
  journal = {Phys. Rev. Lett.},
  volume = {80},
  issue = {24},
  pages = {5243--5246},
  numpages = {0},
  year = {1998},
  month = {Jun},
  publisher = {American Physical Society},
  doi = {10.1103/PhysRevLett.80.5243},
  url = {https://link.aps.org/doi/10.1103/PhysRevLett.80.5243}
}

@article{Hn1,
  title = {Vortex pinning and non-Hermitian quantum mechanics},
  author = {Hatano, Naomichi and Nelson, David R.},
  journal = {Phys. Rev. B},
  volume = {56},
  issue = {14},
  pages = {8651--8673},
  numpages = {0},
  year = {1997},
  month = {Oct},
  publisher = {American Physical Society},
  doi = {10.1103/PhysRevB.56.8651},
  url = {https://link.aps.org/doi/10.1103/PhysRevB.56.8651}
}

@article{PhysRevB.107.L121101,
  title = {Electric polarization and its quantization in one-dimensional non-Hermitian chains},
  author = {Hu, Jinbing and Perroni, Carmine Antonio and De Filippis, Giulio and Zhuang, Songlin and Marrucci, Lorenzo and Cardano, Filippo},
  journal = {Phys. Rev. B},
  volume = {107},
  issue = {12},
  pages = {L121101},
  numpages = {6},
  year = {2023},
  month = {Mar},
  publisher = {American Physical Society},
  doi = {10.1103/PhysRevB.107.L121101},
  url = {https://link.aps.org/doi/10.1103/PhysRevB.107.L121101}
}

@ARTICLE{Wu2022-wx,
  title    = "Enhanced chiroptical responses through coherent perfect
              absorption in a parity-time symmetric system",
  author   = "Wu, Hsin-Yu and Vollmer, Frank",
  journal  = "Commun. Phys.",
  volume   =  5,
  number   =  1,
  pages    = "78",
  month    =  apr,
  year     =  2022,
url = "https://www.nature.com/articles/s42005-022-00855-w"
}

@article{PhysRevLett.112.203901,
  title = {Enhancing the Sensitivity of Frequency and Energy Splitting Detection by Using Exceptional Points: Application to Microcavity Sensors for Single-Particle Detection},
  author = {Wiersig, Jan},
  journal = {Phys. Rev. Lett.},
  volume = {112},
  issue = {20},
  pages = {203901},
  numpages = {5},
  year = {2014},
  month = {May},
  publisher = {American Physical Society},
  doi = {10.1103/PhysRevLett.112.203901},
  url = {https://link.aps.org/doi/10.1103/PhysRevLett.112.203901}
}

@article{FAN2024110434,
title = {Sensitivity enhancement at higher order exceptional point based on coupled whispering gallery modes in microstructured optical fibers},
journal = {Opt. Laser Technol.},
volume = {172},
pages = {110434},
year = {2024},
issn = {0030-3992},
doi = {https://doi.org/10.1016/j.optlastec.2023.110434},
url = {https://www.sciencedirect.com/science/article/pii/S0030399223013270},
author = {Miaosen Fan and Hao Zhang and Xiaoli Shan and Miaoling Yang and Yuan Yao and Wei Lin and Bo Liu}
}

@article{PhysRevLett.123.180501,
  title = {Quantum Noise Theory of Exceptional Point Amplifying Sensors},
  author = {Zhang, Mengzhen and Sweeney, William and Hsu, Chia Wei and Yang, Lan and Stone, A. D. and Jiang, Liang},
  journal = {Phys. Rev. Lett.},
  volume = {123},
  issue = {18},
  pages = {180501},
  numpages = {6},
  year = {2019},
  month = {Oct},
  publisher = {American Physical Society},
  doi = {10.1103/PhysRevLett.123.180501},
  url = {https://link.aps.org/doi/10.1103/PhysRevLett.123.180501}
}

@Article{Zhang2021,
author={Zhang, Xintong
and Xu, Ke
and Liu, Chunmin
and Song, Xiaoxiao
and Hou, Bowen
and Yu, Rui
and Zhang, Hao
and Li, Dan
and Li, Jing},
title={Gauge-dependent topology in non-reciprocal hopping systems with pseudo-Hermitian symmetry},
journal={Commun. Phys.},
year={2021},
month={Jul},
day={20},
volume={4},
number={1},
pages={166},
issn={2399-3650},
doi={10.1038/s42005-021-00668-3},
url={https://doi.org/10.1038/s42005-021-00668-3}
}

@article{HN2,
  title = {Localization Transitions in Non-Hermitian Quantum Mechanics},
  author = {Hatano, Naomichi and Nelson, David R.},
  journal = {Phys. Rev. Lett.},
  volume = {77},
  issue = {3},
  pages = {570--573},
  numpages = {0},
  year = {1996},
  month = {Jul},
  publisher = {American Physical Society},
  doi = {10.1103/PhysRevLett.77.570},
  url = {https://link.aps.org/doi/10.1103/PhysRevLett.77.570}
}

@article{eps-sensors,
author = {Jan Wiersig},
journal = {Photon. Res.},
number = {9},
pages = {1457--1467},
publisher = {Optica Publishing Group},
title = {Review of exceptional point-based sensors},
volume = {8},
month = {Sep},
year = {2020},
url = {https://opg.optica.org/prj/abstract.cfm?URI=prj-8-9-1457},
doi = {10.1364/PRJ.396115},
}

@article{eps-science,
author = {Mohammad-Ali Miri  and Andrea Alù },
title = {Exceptional points in optics and photonics},
journal = {Science},
volume = {363},
number = {6422},
pages = {eaar7709},
year = {2019},
doi = {10.1126/science.aar7709},
URL = {https://www.science.org/doi/abs/10.1126/science.aar7709}}

@article{Longhi-PT,
author = {Stefano Longhi},
journal = {Opt. Lett.},
number = {6},
pages = {1591--1594},
publisher = {Optica Publishing Group},
title = {Quantum statistical signature of {PT} symmetry breaking},
volume = {45},
month = {Mar},
year = {2020},
url = {https://opg.optica.org/ol/abstract.cfm?URI=ol-45-6-1591},
doi = {10.1364/OL.386232},
}

@ARTICLE{Nori-PT,
  title    = "Parity--time symmetry and exceptional points in photonics",
  author   = "{\"O}zdemir, {\c S} K and Rotter, S and Nori, F and Yang, L",
  journal  = "Nat. Mater.",
  volume   =  18,
  number   =  8,
  pages    = "783--798",
  month    =  aug,
  year     =  2019,
url = "https://www.nature.com/articles/s41563-019-0304-9"
}

@article{PhysRevLett.125.186802,
  title = {Non-Hermitian Skin Modes Induced by On-Site Dissipations and Chiral Tunneling Effect},
  author = {Yi, Yifei and Yang, Zhesen},
  journal = {Phys. Rev. Lett.},
  volume = {125},
  issue = {18},
  pages = {186802},
  numpages = {7},
  year = {2020},
  month = {Oct},
  publisher = {American Physical Society},
  doi = {10.1103/PhysRevLett.125.186802},
  url = {https://link.aps.org/doi/10.1103/PhysRevLett.125.186802}
}

\clearpage
{\Large \bf End Matter}
\vspace{.5cm}\\
\vspace{.5cm}\\
 \begin{figure*}[t!]
    \centering
    \includegraphics[width=0.87\linewidth]{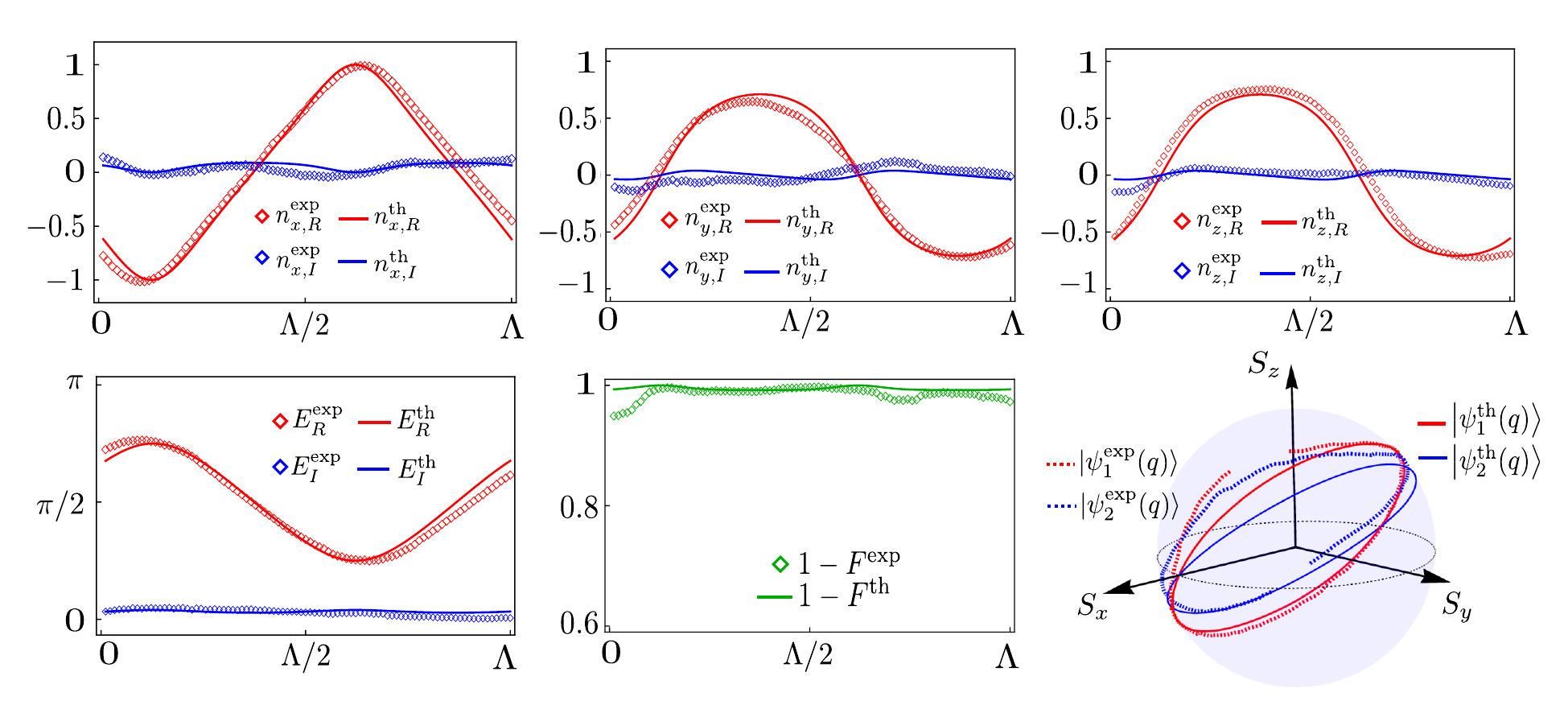}
    \caption{\textbf{Experimental results for a weakly non-Hermitian regime.} Tomographic reconstructions for the case ${(\delta,\eta)=(\pi,0.25)}$ across one spatial period $\Lambda$, corresponding to the first BZ. Real and imaginary parts of the energy bands and the $\textbf{n}$-vector components are extracted and compared with theoretical predictions. The overlap between the right eigenstates is also reported as infidelity at each quasi-momentum, and the trajectories of the polarization eigenstates are visualized on the Bloch sphere.}
    \label{fig:figS4}
\end{figure*}
\textbf{Appendix A: Weakly non-Hermitian regime}
\vspace{.3cm}
\\
Figure \figref{fig:figS4} shows the tomographic reconstructions for the case ${\left(\delta,\eta \right)=(\pi,0.25)}$. Here, the dichroic parameter $\eta$ is relatively low, corresponding to a weakly non-unitary evolution. Accordingly, the imaginary part of all features remains very close to zero, with infidelity close to one at every $q$, corresponding to nearly orthogonal eigenstates. The average fidelity is $\bar{\mathcal{F}}=(98.5\pm 0.8)\%$, computed as in Fig. \figref{fig:fig3} of the main text.
 From the reconstructed $\textbf{n}$ vectors, we obtain $\nu=0.95$, fully compatible with the theory. The non-trivial topology of the energy bands at ${\delta=\pi}$ (see Fig. \figref{fig:fig1}(b)) is also reflected into closed loops without self-intersection points across the BZ, as already observed in Fig. \figref{fig:fig3}(b).

\vspace{.3cm}
\textbf{Appendix B: Sublattice and PT Symmetry in Two-Level Systems}
\vspace{.3cm}
\\
For the sake of completeness, we connect here PT-symmetry and sublattice symmetry for a qubit. We show that, given a non-Hermitian Hamiltonian for which sublattice symmetry holds, its eigenvalues are both purely
real or both purely imaginary if and only if the non-Hermitian Hamiltonian is unitarily equivalent to a PT-symmetric or to an anti-PT-symmetric matrix. We notice that the two eigenvalues are also opposite to each other, which is ensured by the sublattice symmetry. More generally, the most generic sublattice non-Hermitian Hamiltonian can also be related to a matrix that is unitarily equivalent to a PT-symmetric matrix through a complex phase factor.

We start by defining a generic sublattice (in general, non-Hermitian) Hamiltonian,
\begin{equation}
\label{chiral-ham}
H_\text{sls}=
    \begin{pmatrix}
        0& z_a&\\
        z_b& 0&
    \end{pmatrix}\,,
\end{equation}
with $z_a$ and $z_b$ generic complex numbers.
This Hamiltonian has eigenvalues opposite in sign, namely
\begin{eqnarray}
\label{opposite-aval}
    \lambda_1=\sqrt{z_a z_b}
    \,,\quad\lambda_2=-\sqrt{z_a z_b}\,,
\end{eqnarray}
and are, in general, complex numbers.
A matrix 
\begin{equation}
\label{HrPhi}
H_{r}(\phi)=
    \begin{pmatrix}
        0& a e^{i \phi}&\\
        b e^{-i \phi}& 0&
    \end{pmatrix}\,,
\end{equation}
with $a$, $b$, and $\phi$ real numbers (as the one in Eq.\,\eqref{rotHam}), is a particular sublattice Hamiltonian.
It has the property that commutes with the operator $VK$, namely
\begin{eqnarray}
&&VK H_{r}(\phi)- H_{r}(\phi) VK=0\,,
\label{rotPT}
\end{eqnarray}
{with}
\begin{eqnarray}
&&V:=-i U_{\phi}^2\,,
\qquad U_{\phi}:=
\begin{pmatrix}
        e^{i \phi/2}& 0&\\
        0& e^{-i\phi/2}&
    \end{pmatrix}\,,
\end{eqnarray}
and $K$ being the complex conjugation operator. The symmetry disclosed in Eq. \Eqref{rotPT} implies the following properties for the right eigenvectors and eigenvalues.
Consider a right eigenvector $\ket{\lambda}$ corresponding to eigenvalue $\lambda$, we have that
\begin{equation}
H_{r}(\phi) (VK \ket{\lambda})=\lambda^* (VK\ket{\lambda})\,,
\end{equation}
namely $VK\ket{\lambda}$ is the eigenvector corresponding to eigenvalue $\lambda^*$. We have two possibilities: 
(i) $VK \ket{\lambda} \propto \ket{\lambda}$ (unbroken $VK$ symmetry), implying that $\lambda^*=\lambda$, namely is real. In this case, considering also Eq. \Eqref{opposite-aval}, the two eigenvalues of $H_r(\phi)$ are of the form  $\lambda_1=r$ and $\lambda_2=-r$, with $r$ real.
(ii)
$VK \ket{\lambda} \not\propto \ket{\lambda}$, implying that the two right eigenvectors are $\ket{\lambda_1}=\ket{\lambda}$, 
and
$\ket{\lambda_2}=VK\ket{\lambda}$ (broken $VK$ symmetry), corresponding to the two eigenvalues $\lambda_1=\lambda$ and $\lambda_2=\lambda^*$, respectively. In this case, considering also Eq. \Eqref{opposite-aval}, the two eigenvalues are purely imaginary, namely of the form  $\lambda_1=ir$ and $\lambda_2=-ir$, with $r$ real.
Spontaneous symmetry breaking can therefore be witnessed 
by 
\begin{equation}
    1-\vert \bra{\lambda} VK \ket{\lambda}  \vert\,,
\end{equation}
which is zero in the unbroken phase and non-zero in the broken phase (see Fig.\,\ref{fig:fig4}).
We notice that the spectrum is either purely real or purely imaginary, \enquote{pure} for brevity.

Moreover, we notice that a generic sublattice Hamiltonian in the form of Eq. \Eqref{chiral-ham} can always be obtained by multiplying $H_r(\phi)$ by an appropriate phase factor $e^{i\phi'}$, 
in formulas:
\begin{equation}
    H_\text{sls}=e^{i \phi'} H_r(\phi)\,.
\end{equation}
As a consequence, while generically leaving the eigenvectors unchanged,
for the eigenvalues to be \enquote{pure} $e^{i \phi'}$ should be either $1$ or $i$. In the case
$e^{i \phi'}=1$, $H_\text{sls}= H_r(\phi)$, and hence the sublattice Hamiltonian is $VK$ symmetric; while for 
$e^{i \phi'}=i$, $H_{sls}=i H_r(\phi)$, and hence the sublattice Hamiltonian is $VK$ anti-symmetric, which means ${VK H_\text{sls}+ H_\text{sls} VK=0}$.
The rotated Hamiltonian presented in the main text (see Eq. \Eqref{rotHam}) turns out to belong to the first class, i.e., it is $VK$-symmetric.
Figure \figref{fig:pt-appendix} shows the theoretical eigenvector coalescence 
and the corresponding \enquote{pure} spectrum, which 
accompanies the spontaneous symmetry breaking, whose experimental verification has been reported in Fig.\,\ref{fig:fig4}.

Finally, we mention that 
the symmetry $VK$ can be related to the more conventional PT symmetry, $\sigma_x K$, via a unitary transformation $R$, namely,
\begin{equation}
    R_\phi^\dag (\sigma_x K) R_\phi=VK\,,
\end{equation}
with
\begin{eqnarray}
R_\phi=W U_\phi^\dag\,,
\end{eqnarray}
$W$ corresponding exactly to the coin rotation operator appearing in Eq.\,\eqref{eqn:singlestep}.
Equivalently, we can also relate $H_r(\phi)$ to a matrix
\begin{equation}
\label{pt-symm-ham}
H_\text{PT}:=\frac{1}{2}
\begin{pmatrix}
i(b-a)& b+a\\
b+a& -i(b-a)
\end{pmatrix}\,,
\end{equation}
that is PT symmetric, namely ${H_\text{PT} (\sigma_x K)-(\sigma_x K) H_\text{PT}=0}$,
via
$H_\text{PT}=R_\phi H_r(\phi) R_\phi^\dag$\,.
Unitary equivalence between the effective Hamiltonians of Eq. \Eqref{HrPhi} and \eqref{pt-symm-ham}
implies that their spectra are identical, and their eigenvectors differ only by the corresponding unitary transformation.
\begin{figure}[b!]
\centering
\includegraphics[width=.4\textwidth]{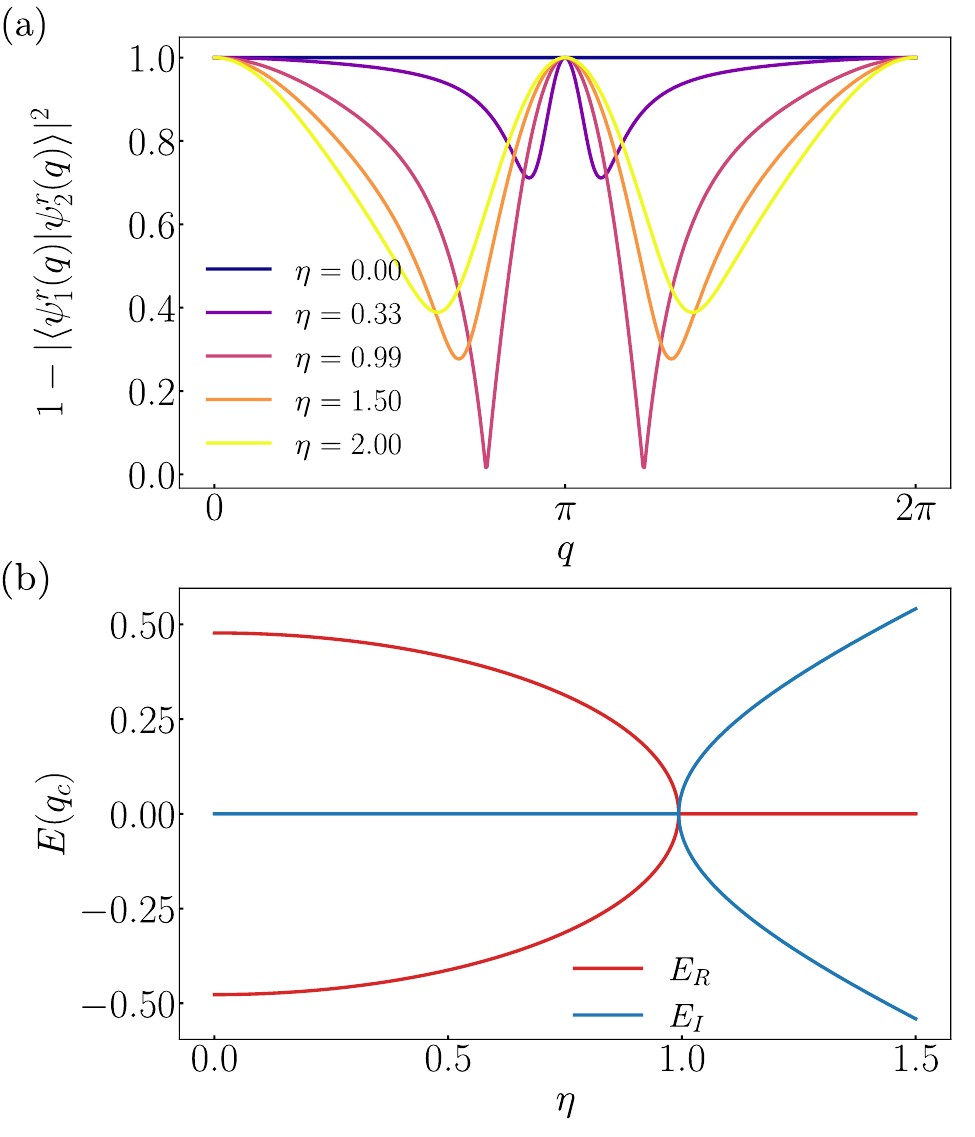}
\caption{\textbf{Exceptional points.} All panels refer to the case ${\delta = 1.3}$, as in Fig.\,\ref{fig:fig4}.
(a)~Theoretical infidelity ${1 - |\langle \psi^r_{1}(q) | \psi^r_{2}(q) \rangle|^2}$. The coalescence of the eigenvectors occurs at $\eta_c \simeq 0.99$ and at the critical quasi-momenta $q_c$ and $2\pi - q_c$, corresponding to the two exceptional points.
(b)~Real and imaginary parts of the two Hamiltonian eigenvalues at the critical quasi-momentum $q_c$, respectively. In the PT-unbroken phase (${\eta < \eta_c}$), the Hamiltonian exhibits real and opposite eigenvalues, while in the PT-broken phase (${\eta > \eta_c}$), the eigenvalues become purely imaginary and opposite. {\color{red}}}
\label{fig:pt-appendix}
\end{figure}

\subsection{Exceptional Points and coalescence of eigenvectors}
We identify the exceptional points following the procedure described in Ref. \Cite{PhysRevA.97.052115}. For a fixed $\delta$, two exceptional points emerge at $q_c$ and ${2\pi - q_c}$ for a critical value of the non-unitary parameter $\eta_c$. At these points, the two eigenvalues simultaneously vanish and the corresponding eigenvectors coalesce.
Figure \ref{fig:pt-appendix}(a) reports the theoretical infidelity, defined as ${1 - F = 1 - |\langle \psi^r_{1}(q) | \psi^r_{2} (q) \rangle|^2}$, between the eigenvectors of the rotated Hamiltonian for ${\delta = 1.3}$ and for several values of $\eta$. The coalescence occurs at the quasi-momentum values $q_c$ and $2\pi - q_c$, at the critical value of the dichroic parameter $\eta_c \simeq 0.99$, consistent with the topological phase transition indicated by the winding number (see Fig. \figref{fig:fig1}(b)).  
As discussed in the main text, in the PT-unbroken (PT-broken) phase the two energy eigenvalues are real (purely imaginary) and opposite, plotted as a red (blue) line in Fig. \figref{fig:pt-appendix}(b). Moreover, in correspondence with the critical value of the dichroic parameter $\eta_c$, one observes that the two eigenvalues vanish simultaneously.

\clearpage
\onecolumngrid
\renewcommand{\figurename}{\textbf{Figure}}
\setcounter{figure}{0} \renewcommand{\thefigure}{\textbf{S{\arabic{figure}}}}
\setcounter{table}{0} \renewcommand{\thetable}{S\arabic{table}}
\setcounter{section}{0} \renewcommand{\thesection}{S\arabic{section}}
\setcounter{equation}{0} \renewcommand{\theequation}{S\arabic{equation}}
\onecolumngrid

\begin{center}
{\Large Supplementary Material for: \\Tomographic characterization of non-Hermitian Hamiltonians in reciprocal space}
\end{center}
\vspace{1 EM}
\section{Experimental setup}

\begin{figure*}[h!]
    \centering
    \includegraphics[width=0.9\linewidth]{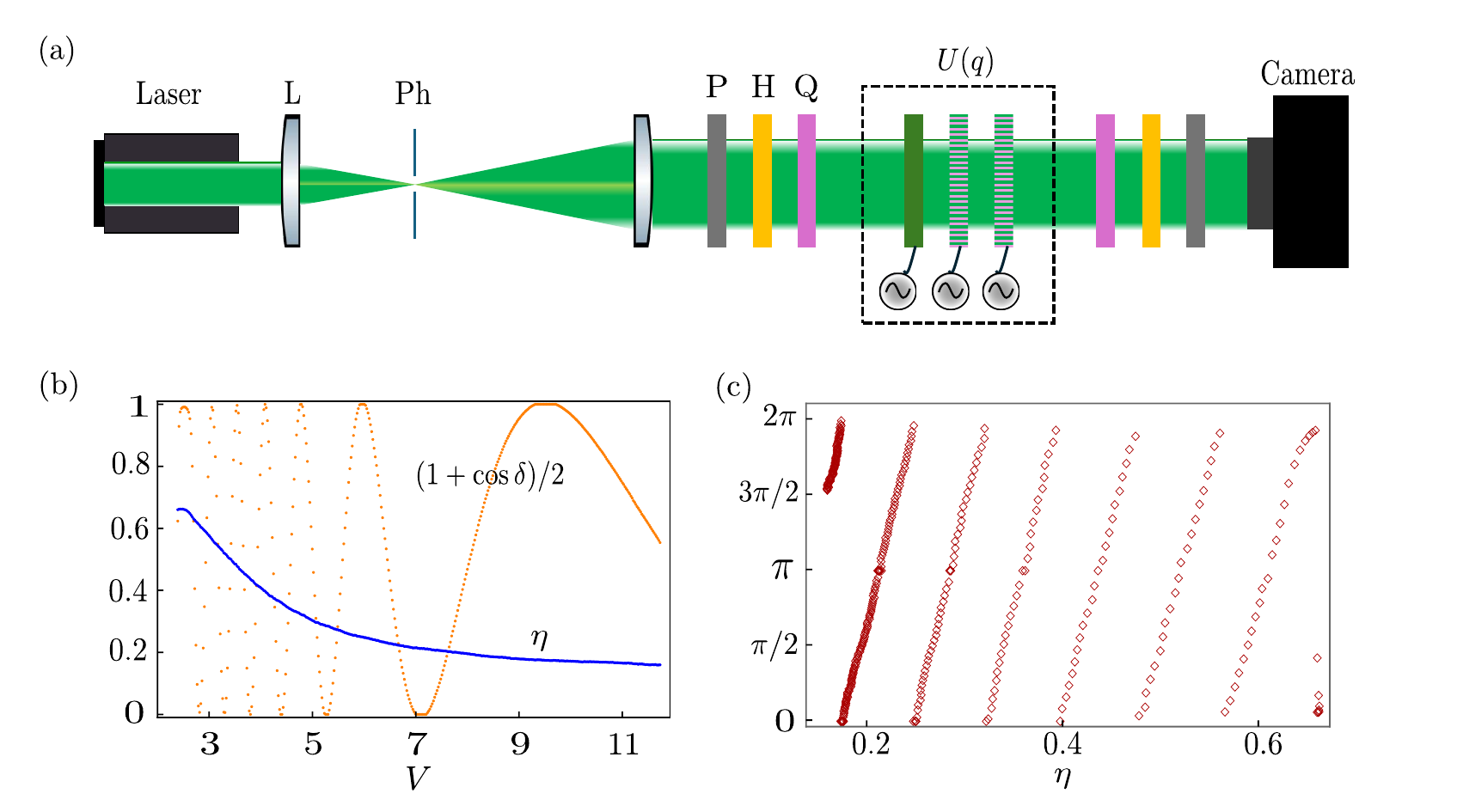}
    \caption{\textbf{Experimental setup.} (a)~A laser beam is expanded with two lenses (L) and spatially filtered with a pinhole (Ph) placed in the focal plane. A uniform, non-dichroic metasurface and two dichroic $g$-plates are employed to simulate a single QW step, consisting of a coin rotation and a non-unitary conditional translation operator. Process tomography is realized by preparing and projecting onto the desired polarization states with a linear polarizer (P), a half-wave plate (H), and a quarter-wave plate (Q). Polarimetric images are processed to retrieve the model eigenstructure. (b)~The application of an external voltage allows for dynamically adjusting the parameters $\delta$ and $\eta$ of dichroic metasurfaces. The electric response of each device is characterized through polarimetric measurements. The indicated voltage values refer to
peak-to-peak amplitudes. (c)~Experimental curve $\delta(\eta)$ obtained for a dichroic metasurface. The periodic character of the birefringence allows for simulating QWs with the same $\delta$ but different values of $\eta$.}
    \label{fig:figS2}
\end{figure*}

The complete experimental setup is depicted in Fig. \figref{fig:figS2}(a). A 532-nm laser beam is expanded via a telescopic configuration of two lenses, having focal lengths ${f_1=35}$ mm and ${f_2=50}$ mm. A 25-$\mu$m pinhole, placed in the focal plane, acts as a spatial filter. The magnified beam covers approximately one BZ on the metasurfaces' plane, which corresponds to 5 mm in our setup. 

The application of an AC voltage allows for dynamically adjusting the parameters $\delta$ and $\eta$ of dichroic metasurfaces, encoding the model topology. The electric response of each device is characterized through polarimetric measurements. 
At a given voltage $V$, the birefringence parameter $\delta$ of a non-dichroic LCMS, such as the one used to implement the coin rotation, is extracted by illuminating the sample with a left-circularly polarized input beam and projecting the transmitted light on the $\ket{\circlearrowleft}$ and $\ket{\circlearrowright}$ polarizations. The value of $\delta$ is obtained as
\begin{equation}
\delta=2\arctan\sqrt\frac{I_{\circlearrowleft\circlearrowright}}{I_{\circlearrowleft\circlearrowleft}},
\label{eqn:delta}
\end{equation}
where ${I_{\circlearrowleft\circlearrowright}}$ (${I_{\circlearrowleft\circlearrowleft}}$) is the intensity of the right (left) projection, corresponding to the converted (unconverted) fraction of incoming light \Cite{Piccirillo2010}.

For dichroic LCMSs, we first extract the dichroic parameter $\eta$ at different voltages. This is accomplished by illuminating a small region of the device having approximately uniform optic-axis orientation. We prepare a small beam with beam waist ${w_0\simeq 50\, \mu m}$ and measure the transmittance of the polarization states aligned with the ordinary and extraordinary axes, from which $\eta$ is obtained as \Cite{savarese1,savarese2}
\begin{equation}
\eta=\frac{1}{2}\log{\frac{I_\text{ord}}{I_\text{ext}}}.
\end{equation}
For extracting $\delta$, we repeat the same procedure as non-dichroic LCMSs, from which we obtain \Cite{savarese1}
\begin{equation}
\delta=\arccos{\left(\frac{I_{\circlearrowleft\circlearrowleft}-I_{\circlearrowleft\circlearrowright}}{I_{\circlearrowleft\circlearrowleft}+I_{\circlearrowleft\circlearrowright}}\,\cosh{\eta}\right)}.
\end{equation}
 
Figure \figref{fig:figS2}(b) illustrates the typical trend of birefringence and dichroic parameters with respect to voltage for one of our fabricated samples. From the data reported in Fig. \figref{fig:figS2}(b), a curve $\delta(\eta)$ can be extracted, as shown in Fig. \figref{fig:figS2}(c).
\newpage
\section{Process tomography routine}
To retrieve the system energy band $E$ and the \textbf{n}-vector components for a given setting of the parameters $\delta$ and $\eta$, we perform 18 polarimetric measurements, realized as prescribed by Eq. \Eqref{eqn:polarimetry}:
\begin{equation}
I_\text{ij}(q)=I_0\abs{\bra{j}U(q)\ket{i}}^2, 
\label{eqn:polarimetryS}
\end{equation}
where $I_0$ is the total light intensity, the input states $\ket{i}$ are extracted from the set ${\lbrace\ket{\circlearrowleft},\ket{H},\ket{D} \rbrace}$ and the projection states $\ket{j}$ are extracted from the set ${\lbrace\ket{\circlearrowleft},\ket{\circlearrowright},\ket{H},\ket{V},\ket{D},\ket{A} \rbrace}$, where $\ket{\circlearrowleft}$ and $\ket{\circlearrowright}$ are the left-handed and right-handed circular polarization states, ${\ket{H}=\left(\ket{\circlearrowleft}+\ket{\circlearrowright}\right)/\sqrt{2}}$ and ${\ket{V}=\left(\ket{\circlearrowleft}-\ket{\circlearrowright}\right)/\sqrt{2}i}$ are the horizontal and vertical polarization states, and ${\ket{D}=\left(\ket{\circlearrowleft}+i\ket{\circlearrowright}\right)/\sqrt{2}}$ and ${\ket{A}=\left(\ket{\circlearrowleft}-i\ket{\circlearrowright}\right)/\sqrt{2}}$ are the diagonal and antidiagonal polarization states, respectively. For each input state $\ket{i}$, the measurement on a given projection $\ket{j}$ is normalized with respect to the orthogonal projection $\ket{j_\perp}$, therefore we obtain

\begin{equation}
\begin{split}
\mathcal{I}_{\circlearrowleft\circlearrowleft}&=\frac{I_{\circlearrowleft\circlearrowleft}}{I_{\circlearrowleft\circlearrowleft}+I_{\circlearrowleft\circlearrowright}};\\
\mathcal{I}_{\circlearrowleft\circlearrowright}&=\frac{I_{\circlearrowleft\circlearrowright}}{I_{\circlearrowleft\circlearrowleft}+I_{\circlearrowleft\circlearrowright}};\\
\mathcal{I}_{\circlearrowleft \text{H}}&=\frac{I_{\circlearrowleft \text{H}}}{I_{\circlearrowleft \text{H}}+I_{\circlearrowleft \text{V}}};\\
\mathcal{I}_{\circlearrowleft \text{V}}&=\frac{I_{\circlearrowleft \text{V}}}{I_{\circlearrowleft \text{H}}+I_{\circlearrowleft \text{V}}};\\
\mathcal{I}_{\circlearrowleft \text{D}}&=\frac{I_{\circlearrowleft \text{D}}}{I_{\circlearrowleft \text{D}}+I_{\circlearrowleft \text{A}}};\\
\mathcal{I}_{\circlearrowleft \text{A}}&=\frac{I_{\circlearrowleft \text{A}}}{I_{\circlearrowleft \text{D}}+I_{\circlearrowleft \text{A}}};\\
\mathcal{I}_{\text{H}\circlearrowleft}&=\frac{I_{\text{H}\circlearrowleft}}{I_{\text{H}\circlearrowleft}+I_{\text{H}\circlearrowright}};\\
\mathcal{I}_{\text{H}\circlearrowright}&=\frac{I_{\text{H}\circlearrowright}}{I_{\text{H}\circlearrowleft}+I_{\text{H}\circlearrowright}};\\
\mathcal{I}_{\text{HH}}&=\frac{I_{\text{HH}}}{I_{\text{HH}}+I_{\text{HV}}};\\
\mathcal{I}_{\text{HV}}&=\frac{I_{\text{HV}}}{I_{\text{HH}}+I_{\text{HV}}};\\
\mathcal{I}_{\text{HD}}&=\frac{I_{\text{HD}}}{I_{\text{HD}}+I_{\text{HA}}};\\
\mathcal{I}_{\text{HA}}&=\frac{I_{\text{HA}}}{I_{\text{HD}}+I_{\text{HA}}};\\
\mathcal{I}_{\text{D}\circlearrowleft}&=\frac{I_{\text{D}\circlearrowleft}}{I_{\text{D}\circlearrowleft}+I_{\text{D}\circlearrowright}};\\
\mathcal{I}_{\text{D}\circlearrowright}&=\frac{I_{\text{D}\circlearrowright}}{I_{\text{D}\circlearrowleft}+I_{\text{D}\circlearrowright}};\\
\mathcal{I}_{\text{DH}}&=\frac{I_{\text{DH}}}{I_{\text{DH}}+I_{\text{DV}}};\\
\mathcal{I}_{\text{DV}}&=\frac{I_{\text{DV}}}{I_{\text{DH}}+I_{\text{DV}}};\\
\mathcal{I}_{\text{DD}}&=\frac{I_{\text{DD}}}{I_{\text{DD}}+I_{\text{DA}}};\\
\mathcal{I}_{\text{DA}}&=\frac{I_{\text{DA}}}{I_{\text{DD}}+I_{\text{DA}}}.
\end{split}
\end{equation}
At each pixel, corresponding to a given quasi-momentum, process tomography is completed by minimizing a cost function $\mathcal{L}$, expressing the distance between the above set of normalized polarimetric measurements and their theoretical expressions as functions of $E=E_R+iE_I$ and $\textbf{n}=\textbf{n}_R+i\textbf{n}_I$, obtained by substituting the decomposition
\begin{equation}
U=\cos{E} \,\sigma_0-i\sin E \left( n_x\sigma_x+n_y\sigma_y+n_z\sigma_z\right)   
\end{equation}
in Eq. \Eqref{eqn:polarimetryS}:
\begin{equation}
\mathcal{L}(q)=\sum_{i,j} \left(\mathcal{I}_{i,j}^\text{exp}(q)-\mathcal{I}_{i,j}^\text{th}(q) \right)^2,
\end{equation}
under the constraint
\begin{equation}
n_x^2+n_y^2+n_z^2=1.
\end{equation}
By separating real and imaginary parts, the latter can be expressed as a double constraint:
\begin{equation}
\begin{split}
n_{xR}^2+n_{yR}^2+n_{zR}^2-n_{xI}^2-n_{yI}^2-n_{zI}^2&=1;\\
n_{xR}n_{xI}+n_{yR}n_{yI}+n_{zR}n_{zI}&=0.
\end{split}
\end{equation}
To facilitate and accelerate convergence, we exploit the continuity of the maps to initialize the numerical minimization routine with the values of $E$ and $\textbf{n}$ obtained for the previous pixel (with the only exception of the first pixel, where the input guess is random). 

\section{Supplementary data}
\begin{figure*}[t!]
    \centering
    \includegraphics[width=0.88\linewidth]{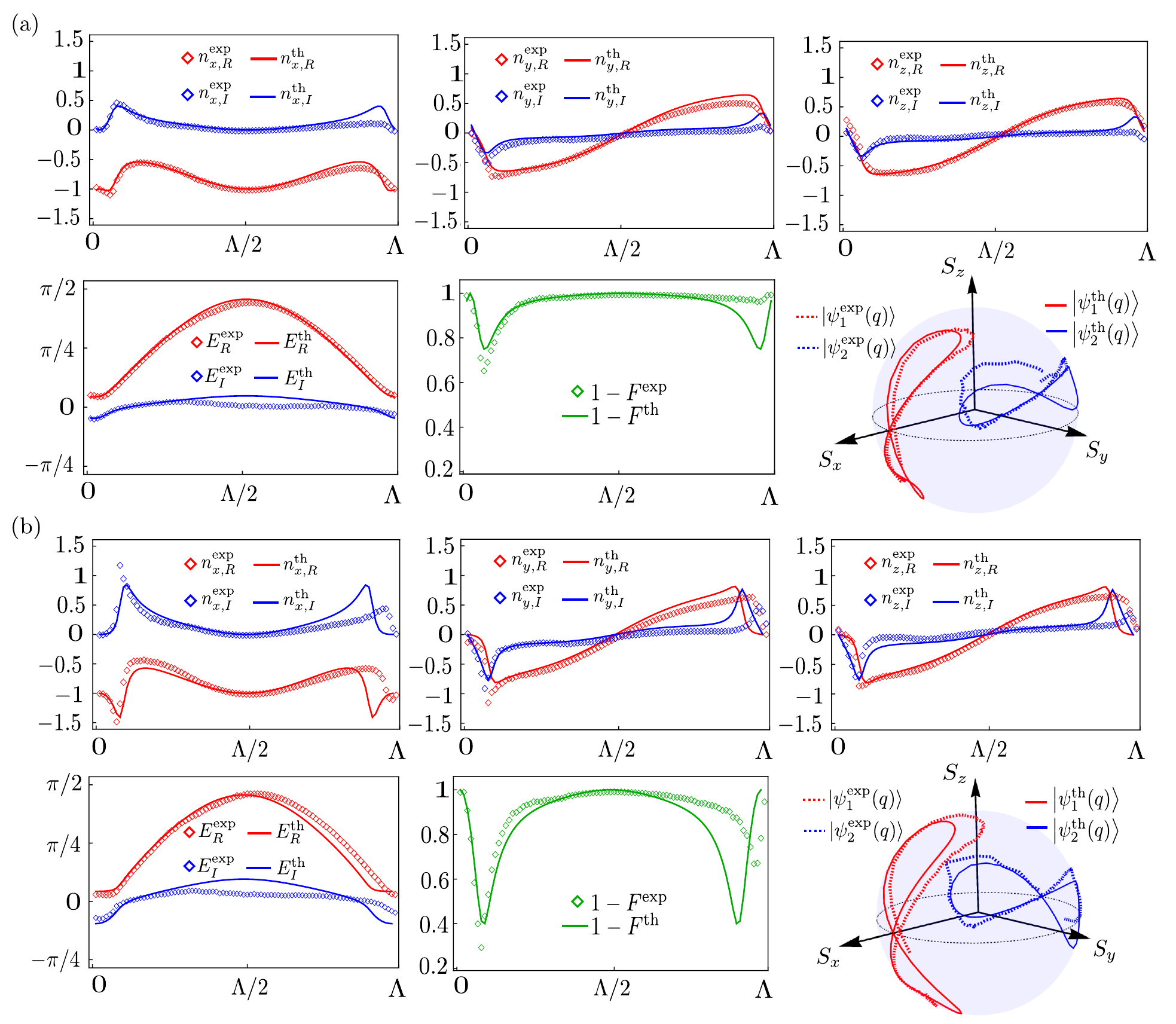}
    \caption{\textbf{Experimental results at ${\delta=1.3}$.} Tomographic reconstructions for the cases: (a)~$(\delta,\eta)=(1.3,0.3)$ and (b)~$(\delta,\eta)=(1.3,0.6)$ across one spatial period $\Lambda$, corresponding to the first BZ. Real and imaginary parts of the energy bands and the $\textbf{n}$-vector components are extracted and compared with theoretical predictions. The overlap between the right eigenstates is also reported as infidelity at each quasi-momentum, and the trajectories of the polarization eigenstates are visualized on the Bloch sphere.}
    \label{fig:figS1}
\end{figure*}

Figure \figref{fig:figS1} reports the tomographic reconstructions of the energy bands and the vector $\textbf{n}$ for the cases ${(\delta,\eta)=(1.3,0.3)}$ and ${(\delta,\eta)=(1.3,0.6)}$. The extracted winding number is approximately 0.02 in both cases, with average fidelities $(99.5\pm0.3)\%$ and $(98.6\pm0.6)\%$, respectively. 

\end{document}